# Performance Evaluation of AODV & DSR with Varying Pause Time & Speed Time Over TCP & CBR Connections in VANET

**Bijan Paul[1], Md.Ibrahim[2] and Md. Abu Naser Bikas[3]**

**[1] Dept. of Computer Science & Engineering, Shahjalal University of Science & Technology**
**Sylhet, Bangladesh**
***bijancse@gmail.com***
**[2] Dept. of Computer Science & Engineering, Shahjalal University of Science & Technology**
**Sylhet, Bangladesh**
***md.ibrahim210@gmail.com***
**[3] Dept. of Computer Science & Engineering, Shahjalal University of Science & Technology**
**Sylhet, Bangladesh**
***bikasbd@yahoo.com***

**Abstract**

VANET (Vehicular Ad-hoc Network) is a new technology which has taken enormous attention in the recent years. Vehicular ad hoc network is formed by cars which are called nodes; allow them to communicate with one another without using any fixed road side unit. It has some unique characteristics which make it different from other ad hoc network as well as difficult to define any exact mobility model and routing protocols because of their high mobility and changing mobility pattern. Hence performance of routing protocols can vary with the various parameters such as speed, pause time, node density and traffic scenarios. In this research paper, the performance of two on-demand routing protocols AODV & DSR has been analyzed by means of packet delivery ratio, loss packet ratio & average end-to-end delay with varying pause time, speed time and node density under TCP & CBR connection.

## 1. Introduction

VANET (Vehicular adhoc network) is an autonomous & self-organizing wireless communication network. In this network the cars are called nodes which involve themselves as servers and/or clients for exchanging & sharing information. This is a new technology thus government has taken huge attention on it. There are many research projects around the world which are related with VANET such as COMCAR [1], DRIVE [2], FleetNet [3] and NoW (Network on Wheels) [4], CarTALK 2000 [5], CarNet [6].

There are several VANET applications such as Vehicle collision warning, Security distance warning, Driver assistance, Cooperative driving, Cooperative cruise control, Dissemination of road information, Internet access, Map location, Automatic parking, and Driverless vehicles.

In this paper, we have evaluated performance of AODV and DSR based on TCP and CBR connection with varying pause time, speed time and also various network parameters and measured performance metrics such as packet delivery ratio, loss packet ratio and average end-to-end delay of this two routing protocol and compared their performance. The remainder of the paper is organized as follows: Section 2 describes previous work related to performance evaluation of AODV and DSR and section 3 discusses about two unicast routing protocols AODV and DSR of VANET. Section 4 describes connection types like TCP and CBR. Section 5 presents performance metrics and the network parameters. Section 6 presents our implementation. Section 7 presents our decisions. We conclude in Section 8 and at the end add references.

## 2.RELATED WORK

There are several papers [7, 8, 9, 10, 11] related to performance evaluation of AODV and DSR .In [7], they measured packet delivery ratio, loss packet ratio and routing overhead using constant speed and pause time. Packet delivery ratio has been measured using variable speed and pause time in [8] .More broaden evaluation has been done in [9].The authors used TCP and UDP connection for performance comparison. Though there is a significant difference between TCP and CBR connection but comparison in between them is not yet analyzed. So we have focused on this two connection pattern based on different. We measured the performance of AODV and DSR with varying speed and constant pause time in [10].We used both high and low node density and observed the performance differences between TCP and CBR connection. Then we used varying pause time and constant speed in [11].We have observed that by changing the speed and pause time the performance varies between two connection. In this paper we have observed and analyzed the performance of AODV and DSR with varying pause time and speed.

## 3. Routing Protocols

An ad hoc routing protocol [12] is a convention, or standard, that controls how nodes decide which way to route packets in between computing devices in a mobile adhoc network.

There are two categories of routing protocol in VANET such as Topology based routing protocols & Position based routing protocols. Existing unicast routing protocols of VANET is not capable to meet every traffic scenarios. They have some pros and cons. We have already described it in our previous work [13]. We have selected two on demand routing protocols AODV & DSR for our simulation purpose.

### 3.1 AODV

Ad Hoc on Demand Distance Vector routing protocol [14] is a reactive routing protocol which establish a route when a node requires sending data packets. It has the ability of unicast & multicast routing. It uses a destination sequence number (DestSeqNum) which makes it different from other on demand routing protocols. It maintains routing tables, one entry per destination and an entry is discarded if it is not used recently. It establishes route by using RREQ and RREP cycle. If any link failure occurs, it sends report and another RREQ is made.

### 3.2 DSR

The Dynamic Source Routing (DSR) [15] protocol utilizes source routing & maintains active routes. It has two phases route discovery & route maintenance. It does not use periodic routing message. It will generate an error message if there is any link failure. All the intermediate nodes ID are stored in the packet header of DSR. If there has multiple paths to go to the destination DSR stores multiple path of its routing information.

AODV and DSR have some significant differences. In AODV when a node sends a packet to the destination then data packets only contains destination address. On the other hand in DSR when a node sends a packet to the destination the full routing information is carried by data packets which causes more routing overhead than AODV.

## 4. CONNECTION TYPES

There are several types of connection pattern in VANET. For our simulation purpose we have used CBR and TCP connection pattern.

### 4.1 Constant Bit Rate (CBR)

Constant bit rate means consistent bits rate in traffic are supplied to the network. In CBR, data packets are sent with fixed size and fixed interval between each data packets. Establishment phase of connection between nodes is not required here, even the receiving node don’t send any acknowledgement messages. Connection is one way direction like source to destination.

### 4.2 Transmission Control Protocol (TCP)

TCP is a connection oriented and reliable transport protocol. To ensure reliable data transfer TCP uses acknowledgement, time outs and retransmission. Acknowledge means successful transmission of packets from source to destination. If an acknowledgement is not received during a certain period of time which is called time out then TCP transmit the data again.

## 5. PERFORMANCE METRICS & NETWORK PARAMETERS

For network simulation, there are several performance metrics which is used to evaluate the performance. In simulation purpose we have used three performance metrics.

### 5.1 Packet Delivery Ratio

Packet delivery ratio is the ratio of number of packets received at the destination to the number of packets sent from the source. The performance is better when packet delivery ratio is high.

### 5.2 Average end-to-end delay

This is the average time delay for data packets from the source node to the destination node. To find out the end-to-end delay the difference of packet sent and received time was stored and then dividing the total time difference over the total number of packet received gave the average end-to-end delay for the received packets. The performance is better when packet end-to-end delay is low.

### 5.3 Loss Packet Ratio (LPR)

Loss Packet Ratio is the ratio of the number of packets that never reached the destination to the number of packets originated by the source.

## 6. OUR IMPLEMENTATION

For simulation purpose we used random waypoint mobility model. Network Simulator NS-2.34[16, 17] has been used. To measure the performance of AODV and DSR we used same scenario for both protocols.

### 6.1 Simulation Parameters

In our simulation, we used environment size 840 m x 840 m, node density 30 to 150 nodes with constant maximum speed 15 m/s and variable pause time 50 to 250 s. We did the Simulation for 200s with maximum 8 connections. The network parameters we have used for our simulation purpose shown in the table 1.

Table 1: Network Parameters

| ***Parameter*** | ***Value*** |
|---|---|
| Protocols | AODV, DSR |
| Simulation Time | 200 s |
| Number of Nodes | 30, 90, 150 |
| Simulation Area | 840 m x 840 m |
| Speed Time | 5 , 10 , 15 , 20 , 25 m / s |
| Pause Time | 50 ,100, 150, 200, 250 s |
| Traffic Type | CBR , TCP |
| Mobility Model | Random Waypoint |
| Network Simulator | NS 2.34 |

## 6.2 Performance Measurement Script

Generally, in NS-2 when we execute a program there creates two types of file trace file and nam file where nam file is used to visualize the simulation and trace file keep records of various interesting quantities such as each individual packets as its arrives, departs or is dropped at a link or queue by which we can measure a protocol performance.

**Trace file format**

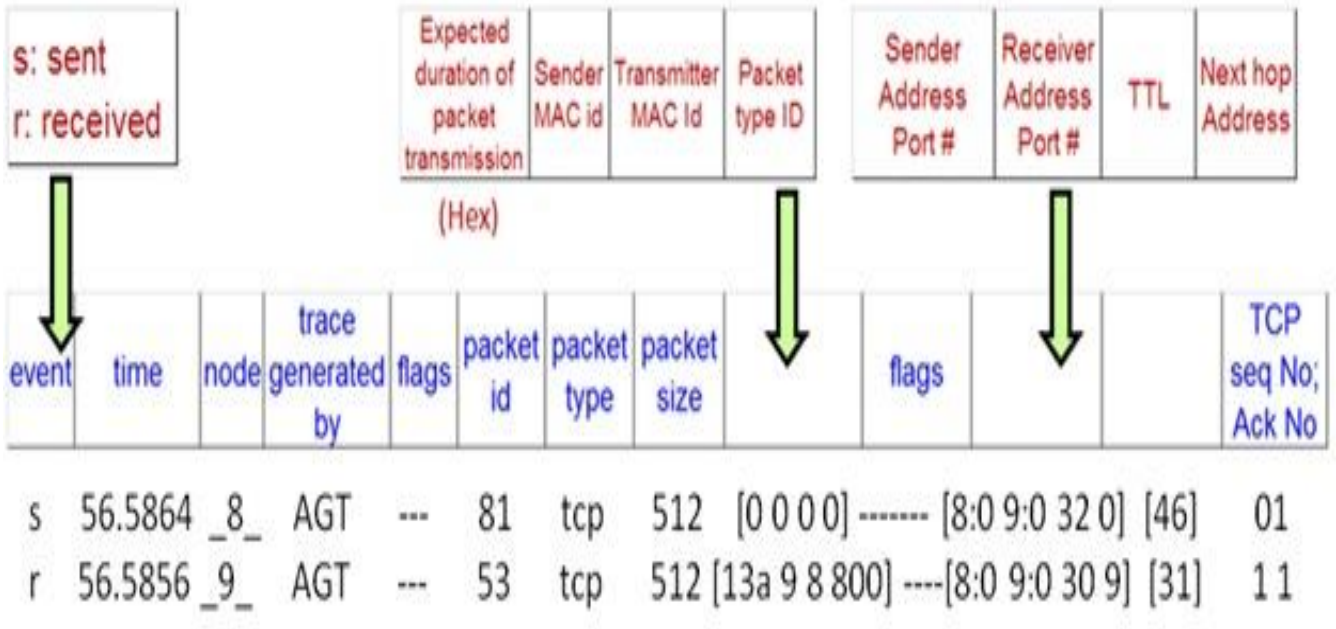

Awk script is required to analysis trace file for performance measure. To measure packet delivery ratio, loss packet ratio & average end-to-end delay of AODV and DSR we make two awk scripts. The scripts sudo codes are given below.

**PDR and LPR Measurement AWK Script**

```
START

//initialization
SET nSentPackets to 0
SET nReceivedPackets to 0

IF $1 =  "s" AND  $4 = "AGT" THEN
    INCREMENT nSentPackets
ENDIF

IF $1 =  "r" AND  $4 = "AGT" THEN
    INCREMENT nReceivedPackets
ENDIF

COMPUTE rPacketDeliveryRatio as nReceivedPackets / nSentPackets * 100

COMPUTE  lpr as ( (nSentPackets-nReceivedPackets) / nSentPackets ) * 100

PRINT  nSentPackets
PRINT  nReceivedPackets
PRINT  rPacketDeliveryRatio
PRINT  lpr
END
```

**END-TO-END DELAY AWK Script**

```
START

//initialization
SET seqno to -1
SET count to 0

IF $4 = "AGT" AND $1 = "s" AND seqno < $6 THEN
    COMPUTE seqno as $6
ENDIF

IF $4 = "AGT" AND  $1 == "s" THEN
    COMPUTE  start_time[$6] as $2
ELSE IF $7 = "tcp" AND $1 = "r" THEN
    COMPUTE  end_time[$6]  as $2
ELSE IF $1 = "D" AND $7 = "tcp" THEN
    COMPUTE  end_time[$6] as -1
ENDIF

FOR   X = 1 to seqno
   IF  end_time[X] > 0 THEN
    COMPUTE delay[X] as end_time[X] - start_time[X]
            INCREMENT count
   ELSE
     COMPUTE  delay[i] as  -1
         ENDIF
END FOR

FOR   X = 1 to seqno
 IF  delay[X] > 0 THEN
  COMPUTE n_to_n_delay as n_to_n_delay + delay[X]
 ENDIF
END FOR

COMPUTE   n_to_n_delay as  n_to_n_delay / count * 1000

PRINT   n_to_n_delay

END
```

## 6.3 Simulation Results Analysis

The performance of AODV & DSR has been analyzed with varying pause time 50s to 250s and speed time 5 to 25 m/s for number of nodes 30, 90, 150 under TCP & CBR connection. We measure the packet delivery ratio, loss packet ratio & average end-to-end delay of AODV and DSR and the simulated output has shown by using graphs.

## 6.4 Graphs

Based on the simulation result we have generated the graph which shows the differences between AODV and DSR. The graphs are given below.

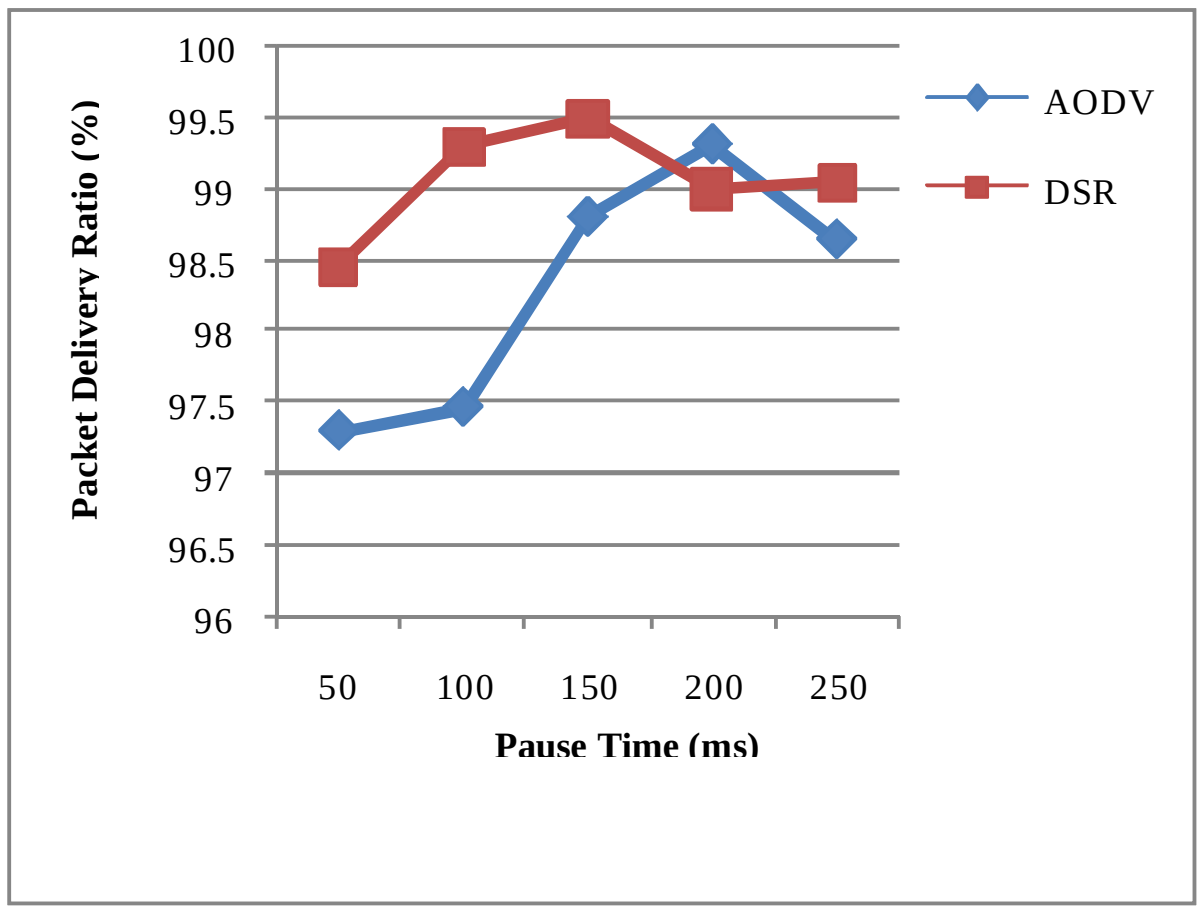

Figure1. PDR (w.r.t. Pause) of 30 nodes using TCP

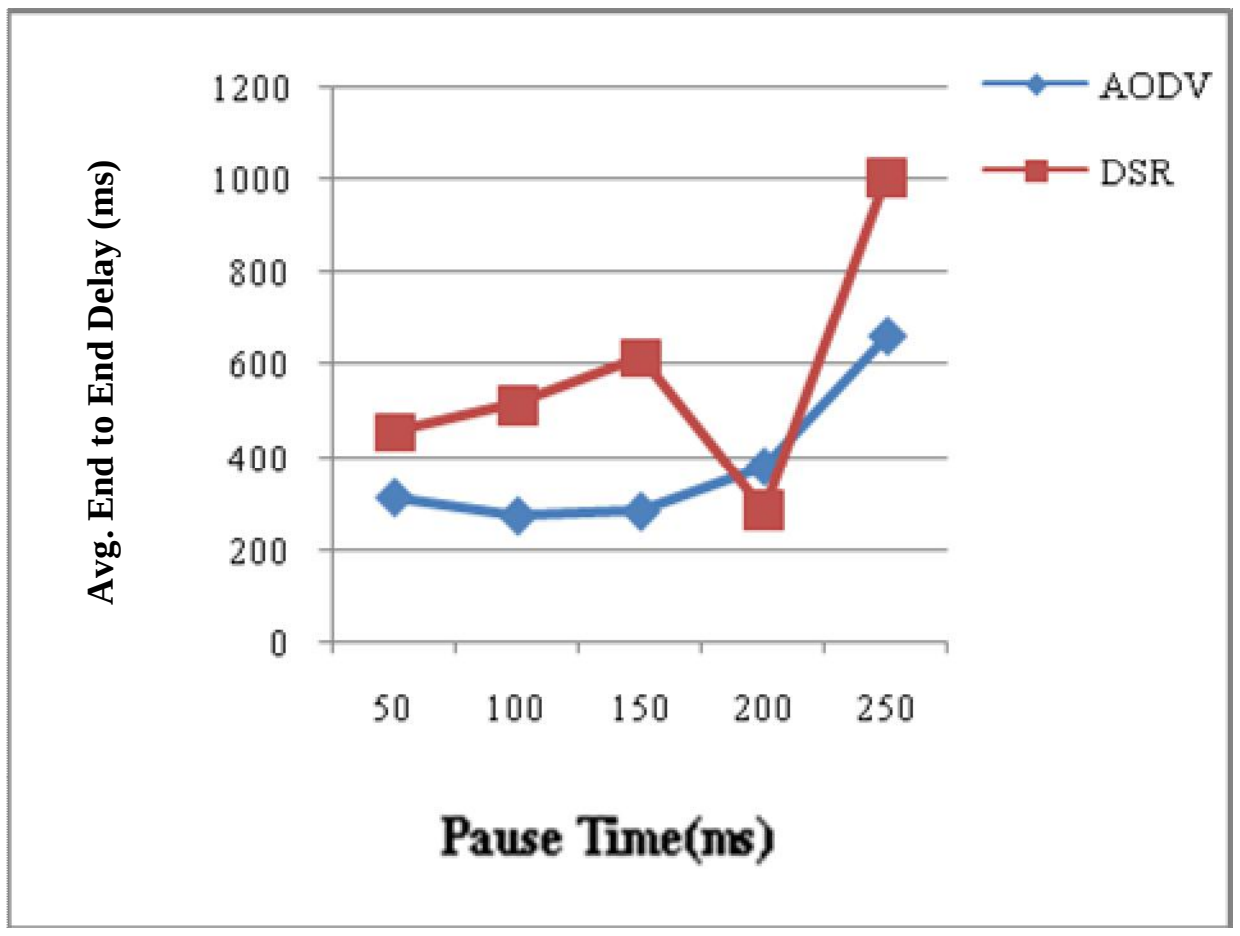

Figure 2. Avg.E2E delay (w.r.t. Pause) of 30 nodes using TCP

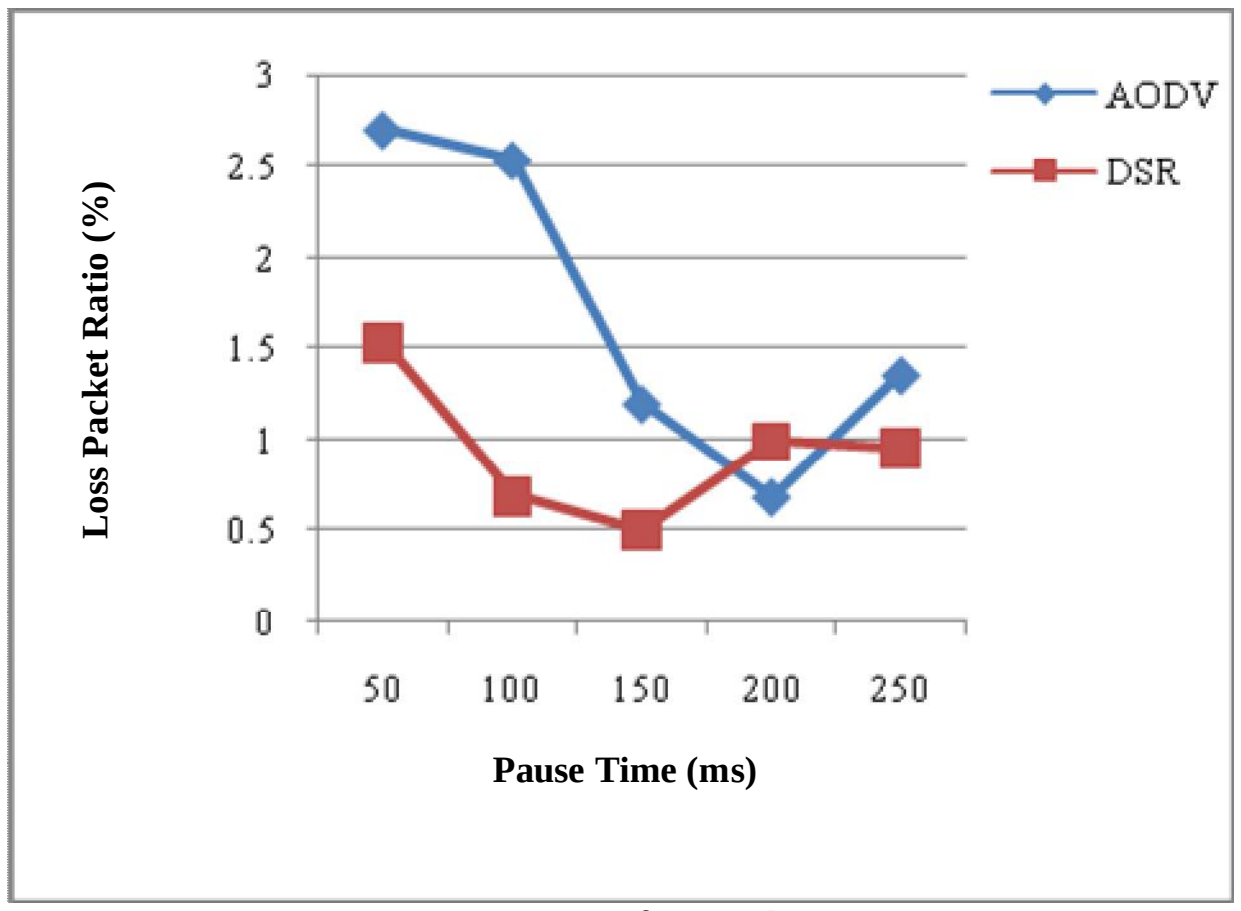

Figure 3. LPR (w.r.t. Pause) of 30 nodes using TCP

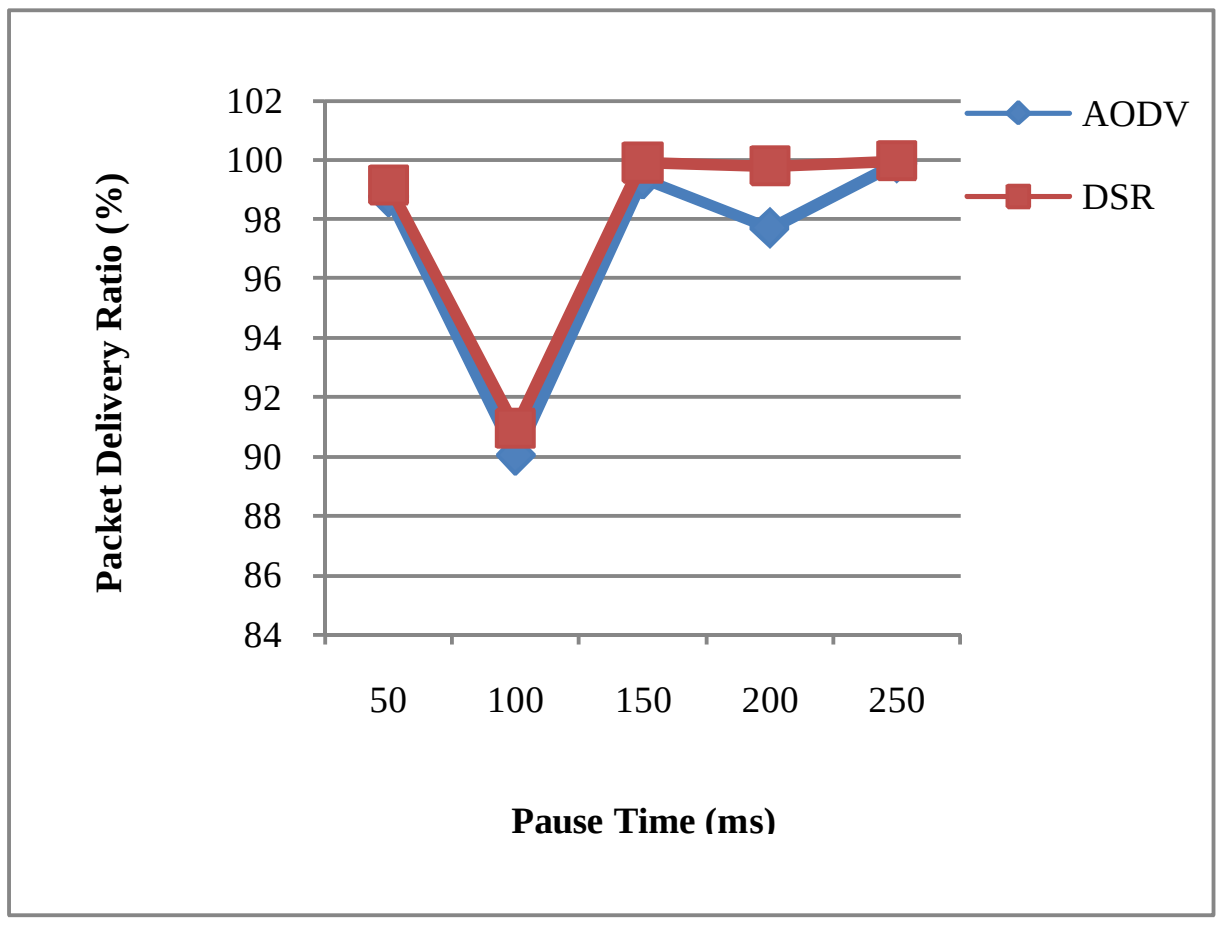

Figure 4. PDR (w.r.t. Pause) of 30 nodes using CBR

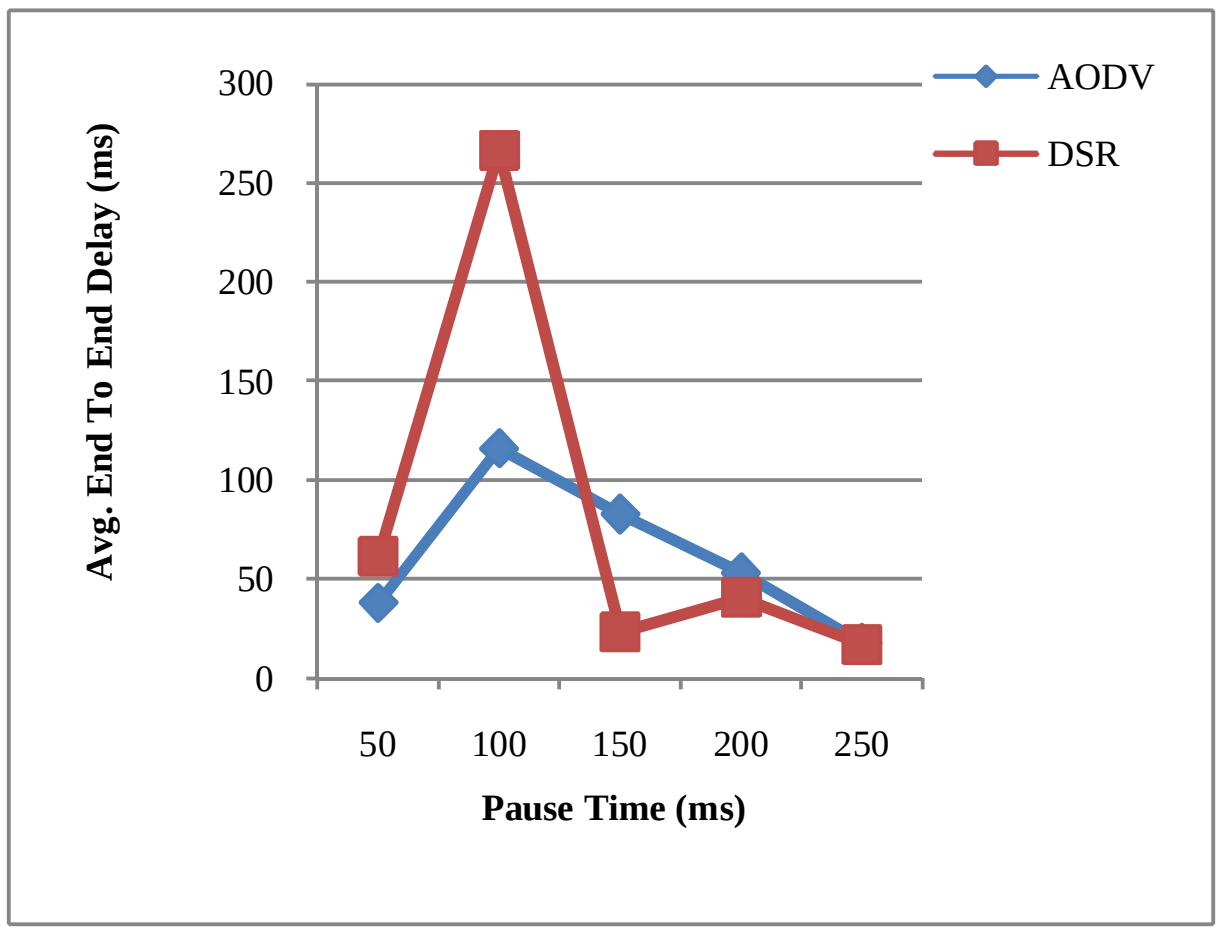

Figure 5. Avg.E2E delay(w.r.t. Pause) of 30 nodes using CBR

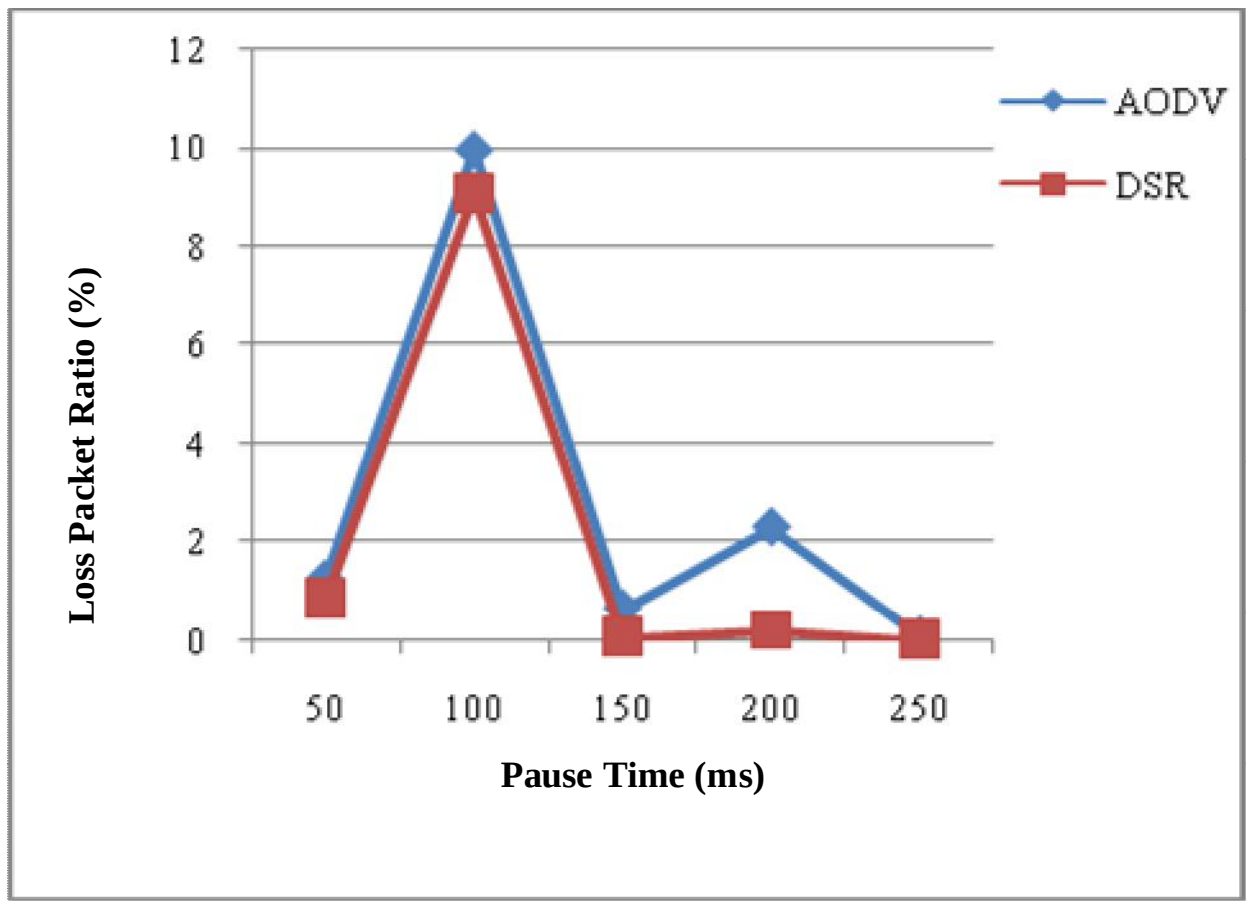

Figure 6. LPR (w.r.t. Pause) of 30 nodes using CBR

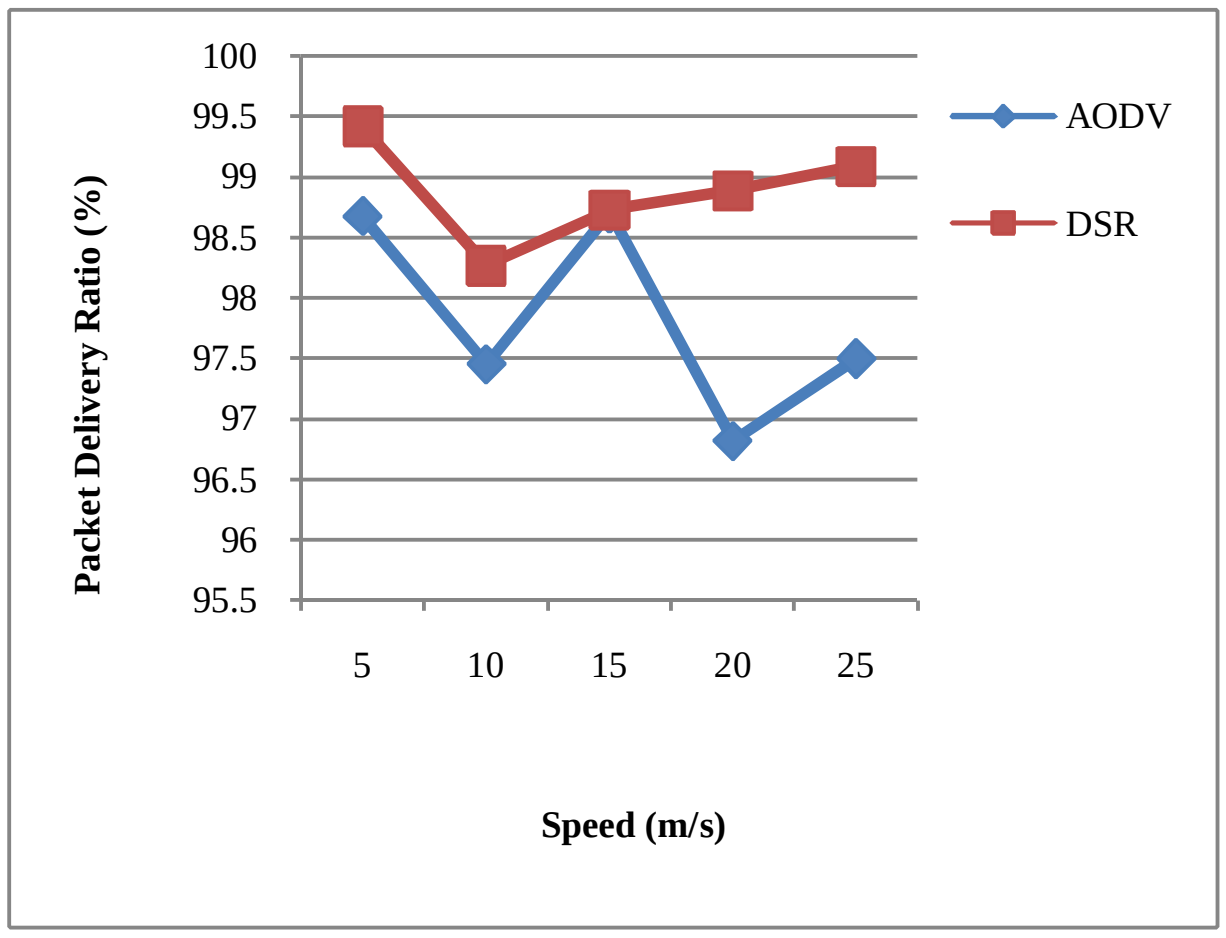

Figure 7. PDR (w.r.t. Speed) of 30 nodes using TCP

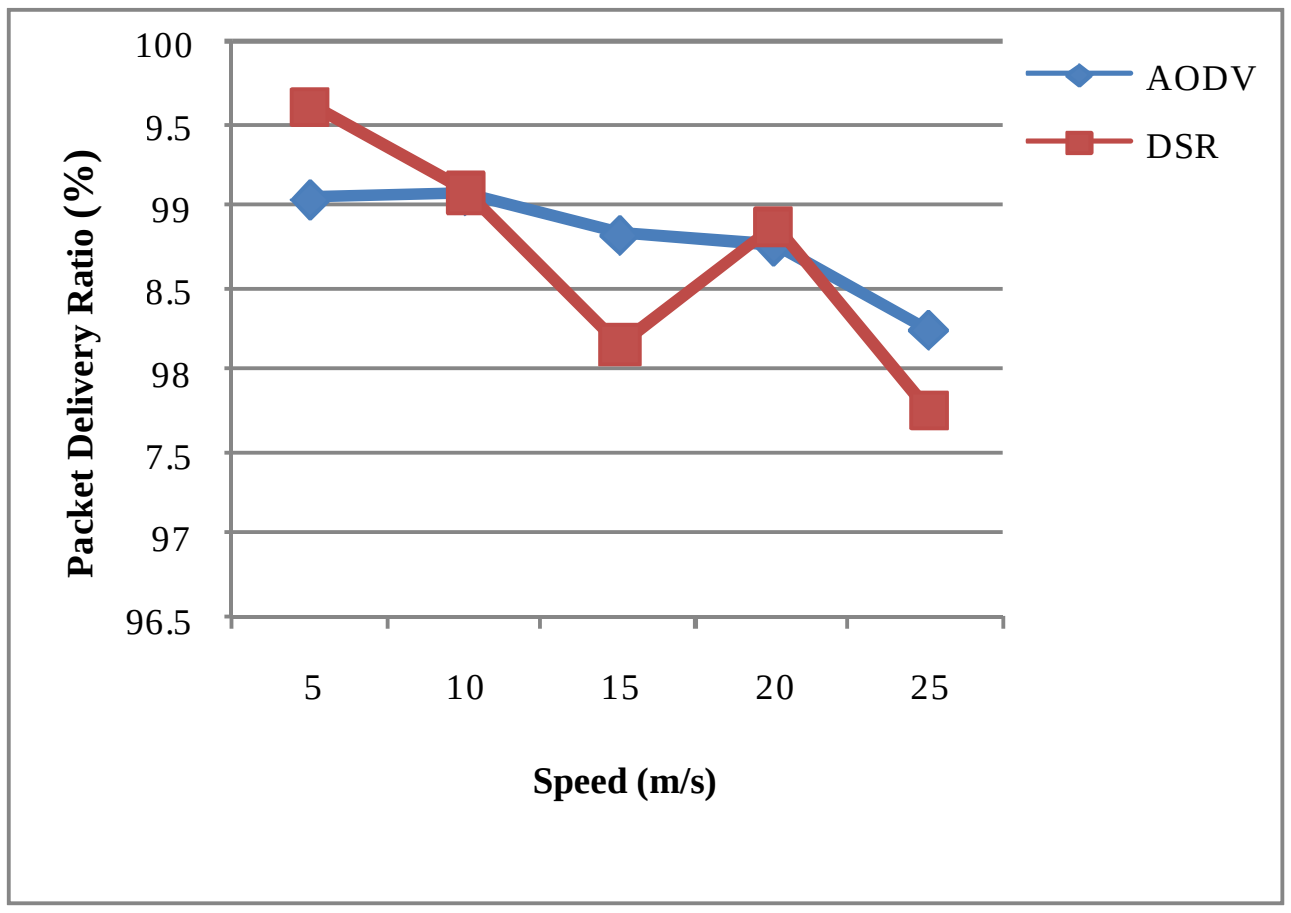

Figure 10. PDR (w.r.t. Speed) of 30 nodes using CBR

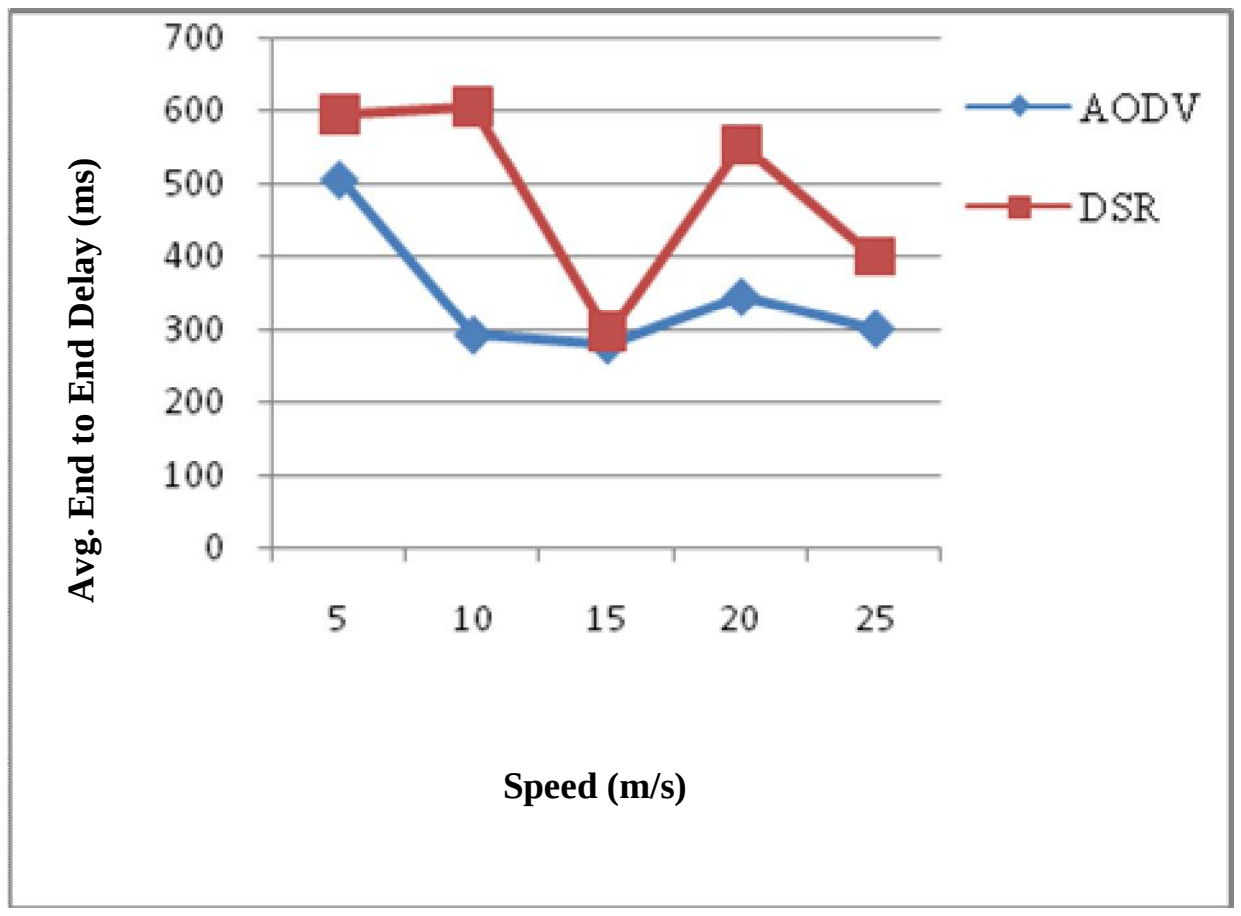

Figure 8. Avg.E2E delay (w.r.t. Speed) of 30 nodes using TCP

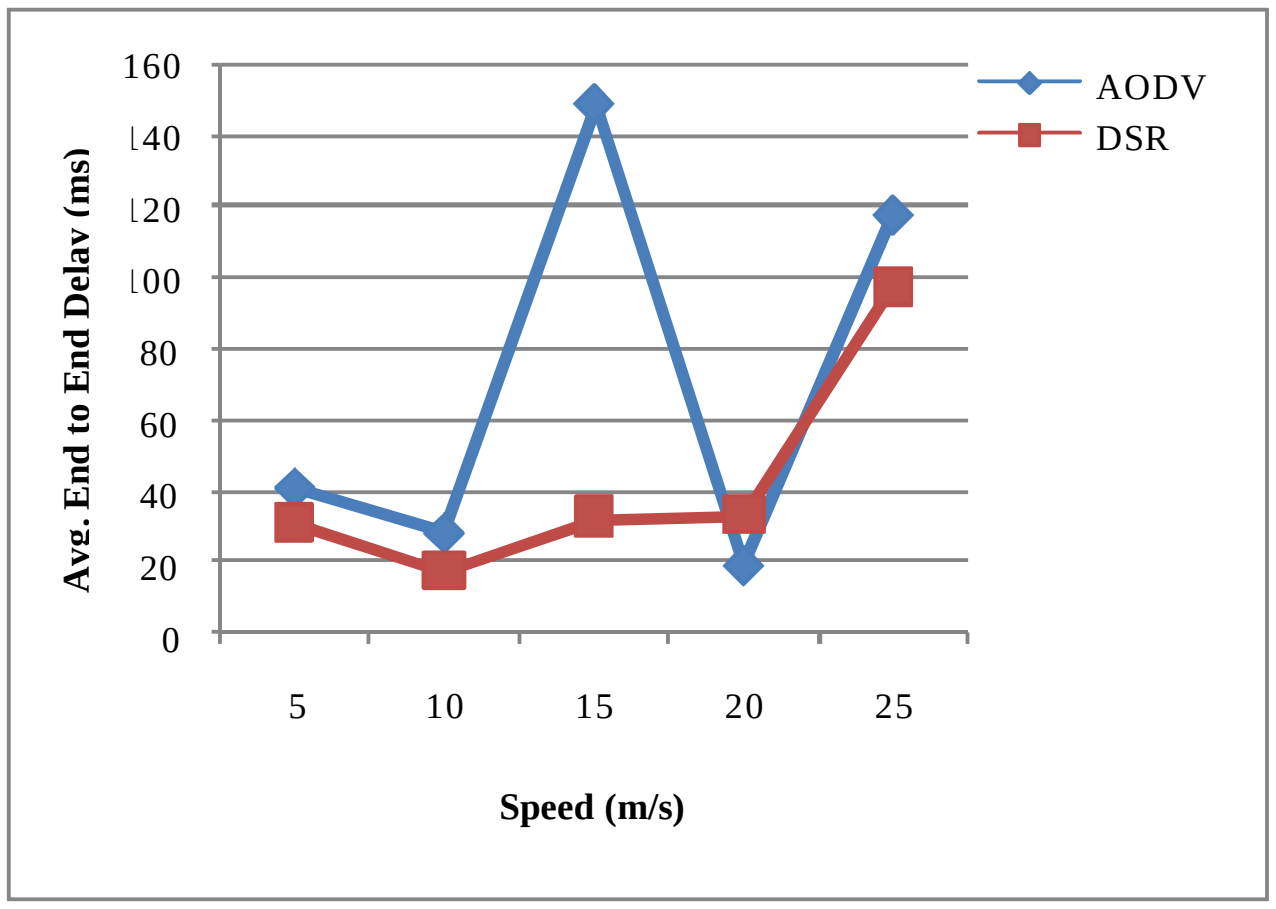

Figure11. Avg.E2E delay (w.r.t. Speed) of 30 nodes using CBR

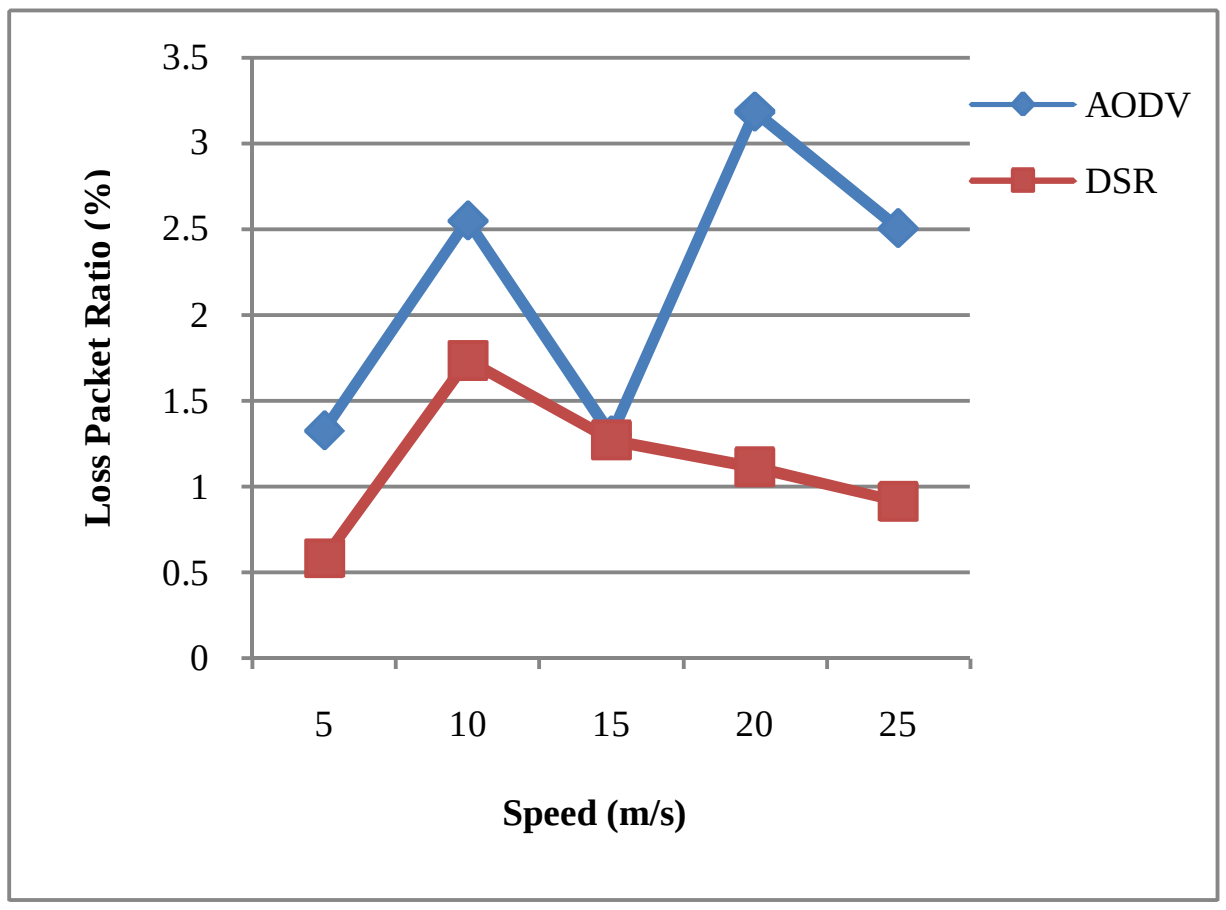

Figure 9. LPR (w.r.t. Speed) of 30 nodes using TCP

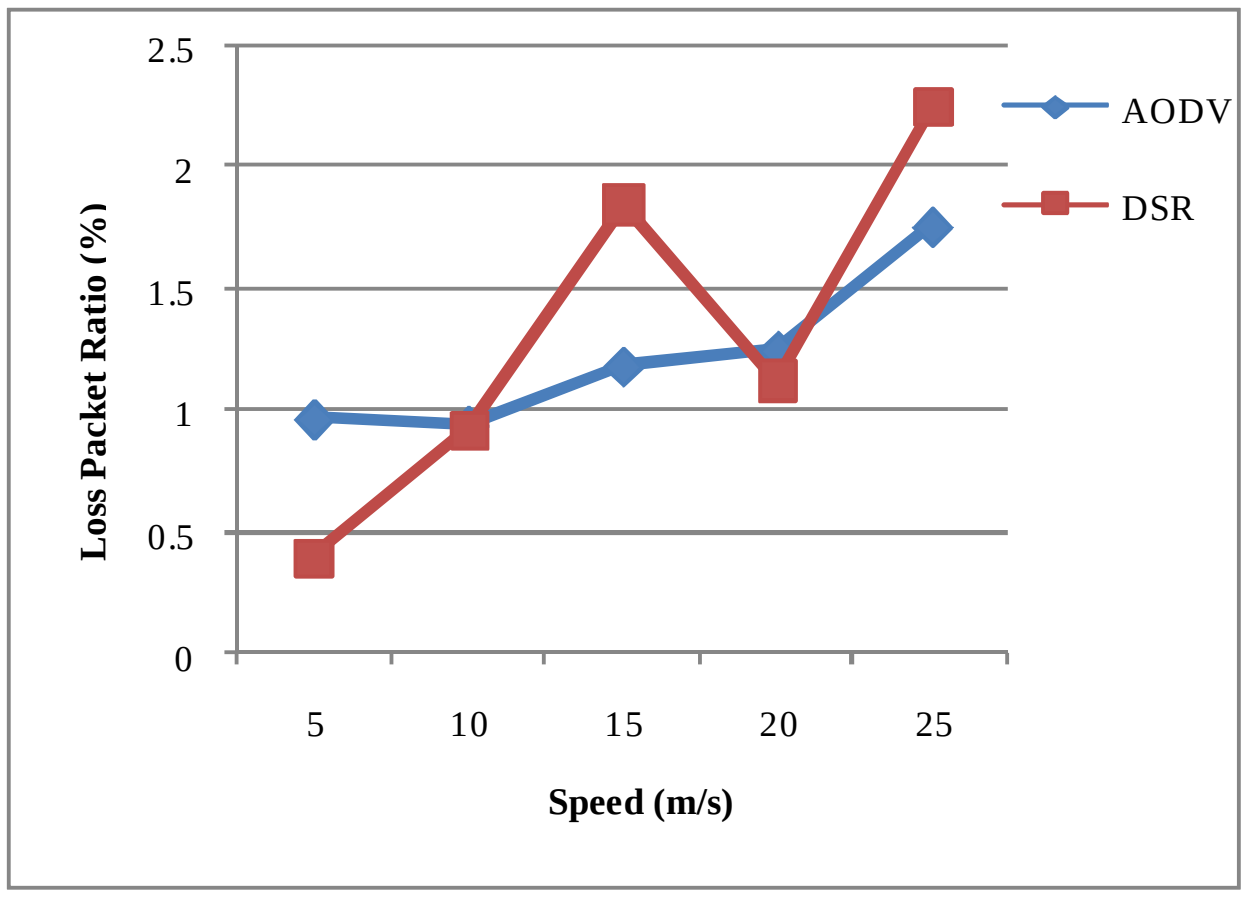

Figure 12. LPR (w.r.t. Speed) of 30 nodes using CBR

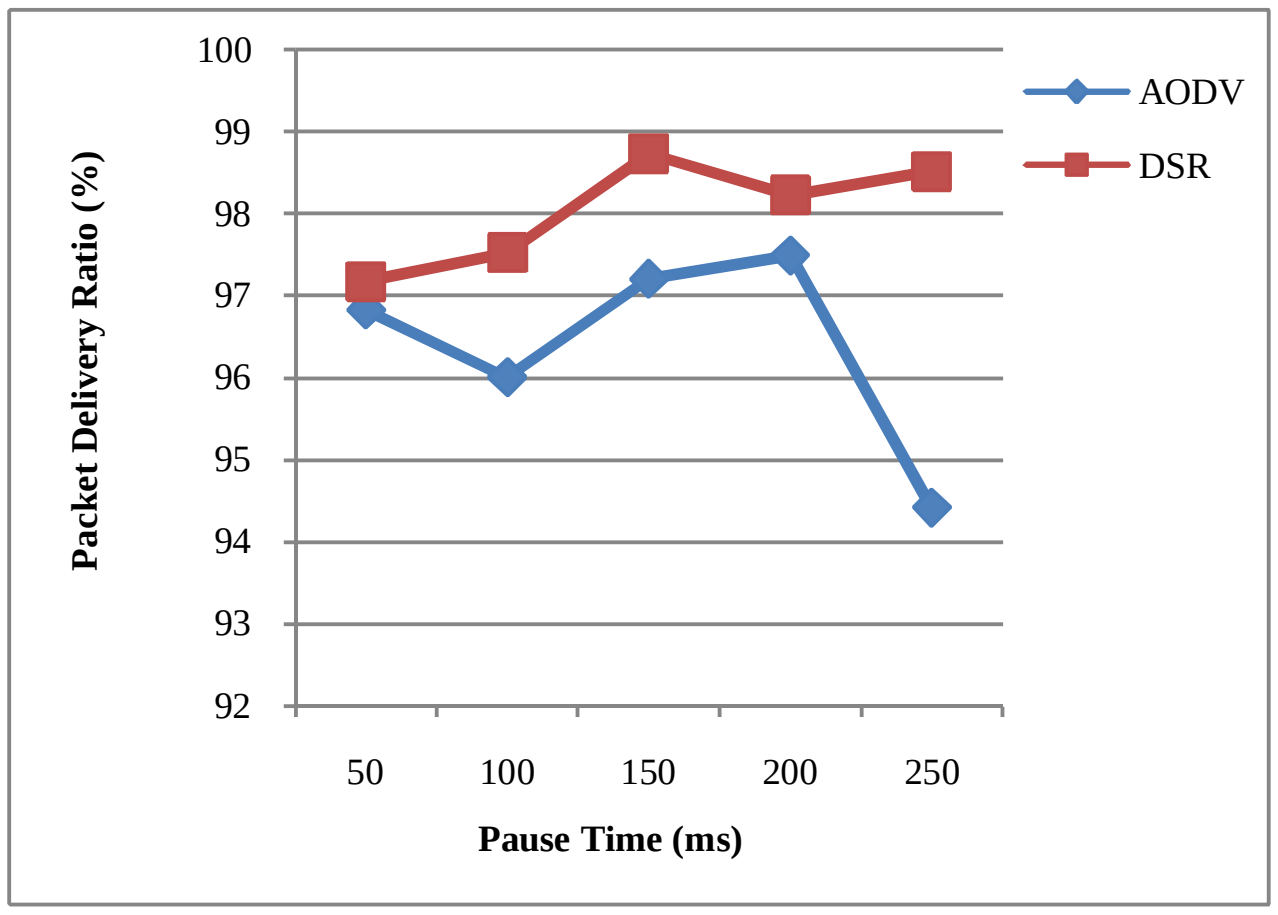

Figure 13. PDR (w.r.t. Pause) of 90 nodes using TCP

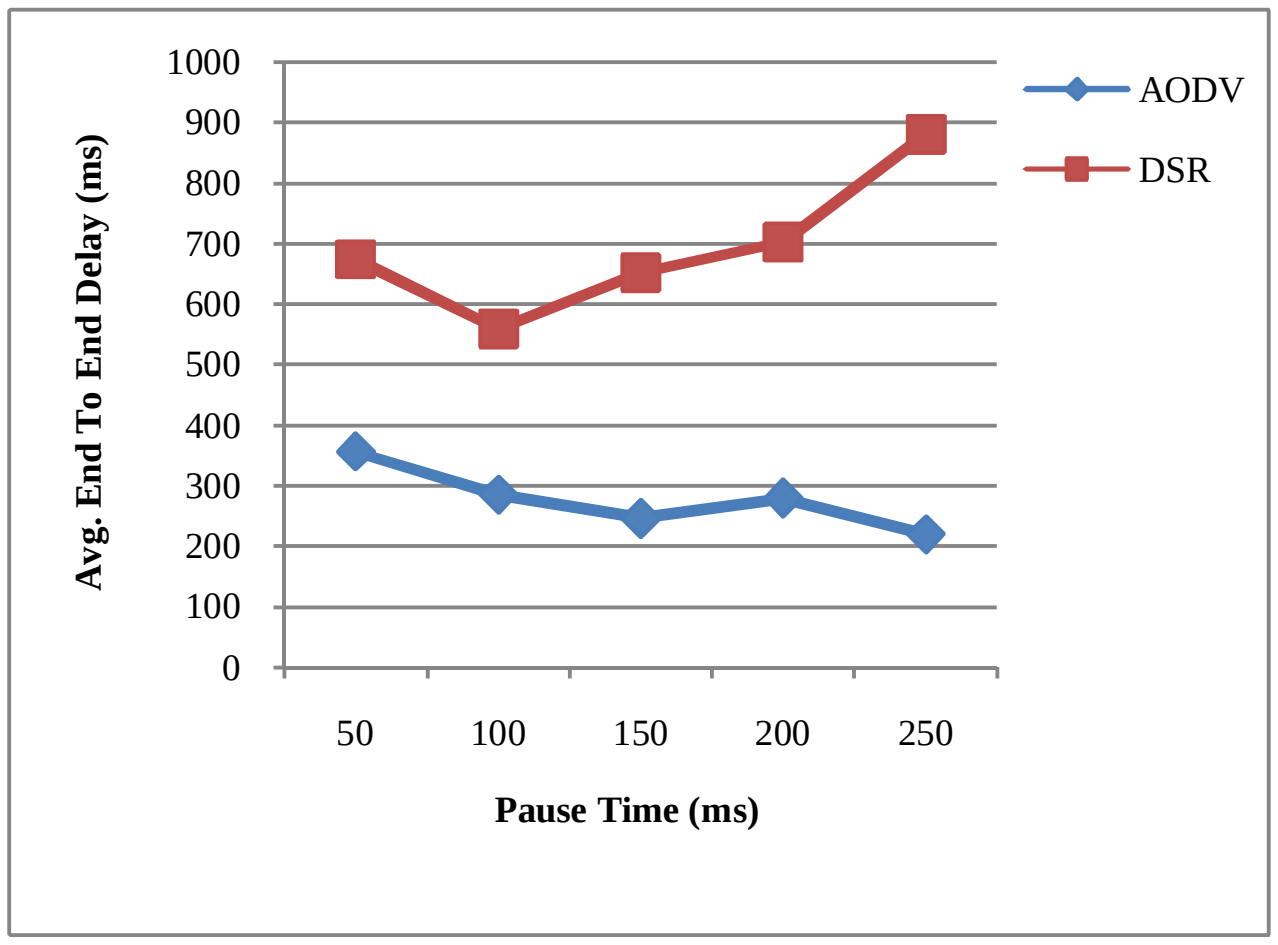

Figure 14. Avg.E-2-E delay (w.r.t. Pause) of 90 nodes using TCP

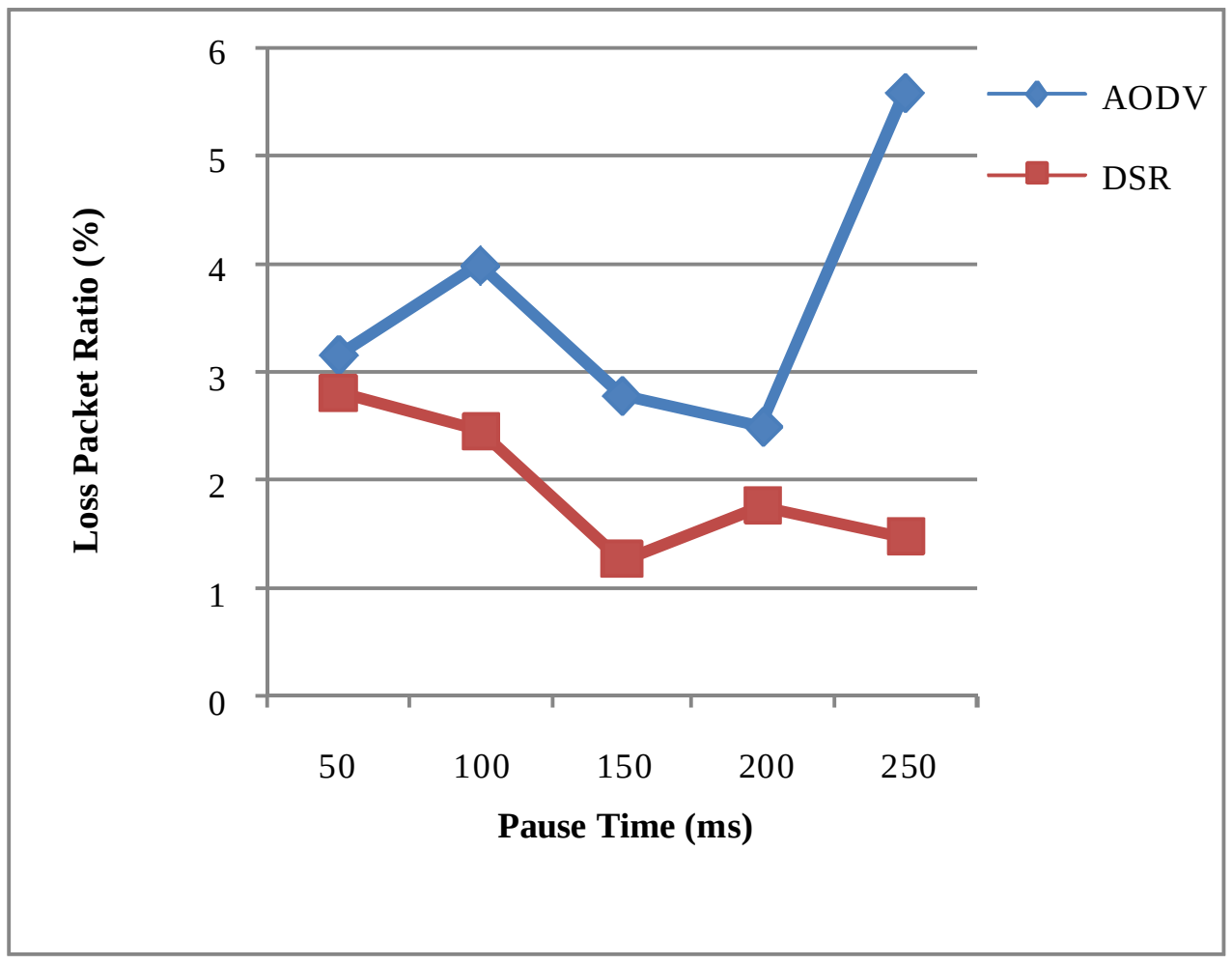

Figure 15. LPR (w.r.t. Pause) of 90 nodes using TCP

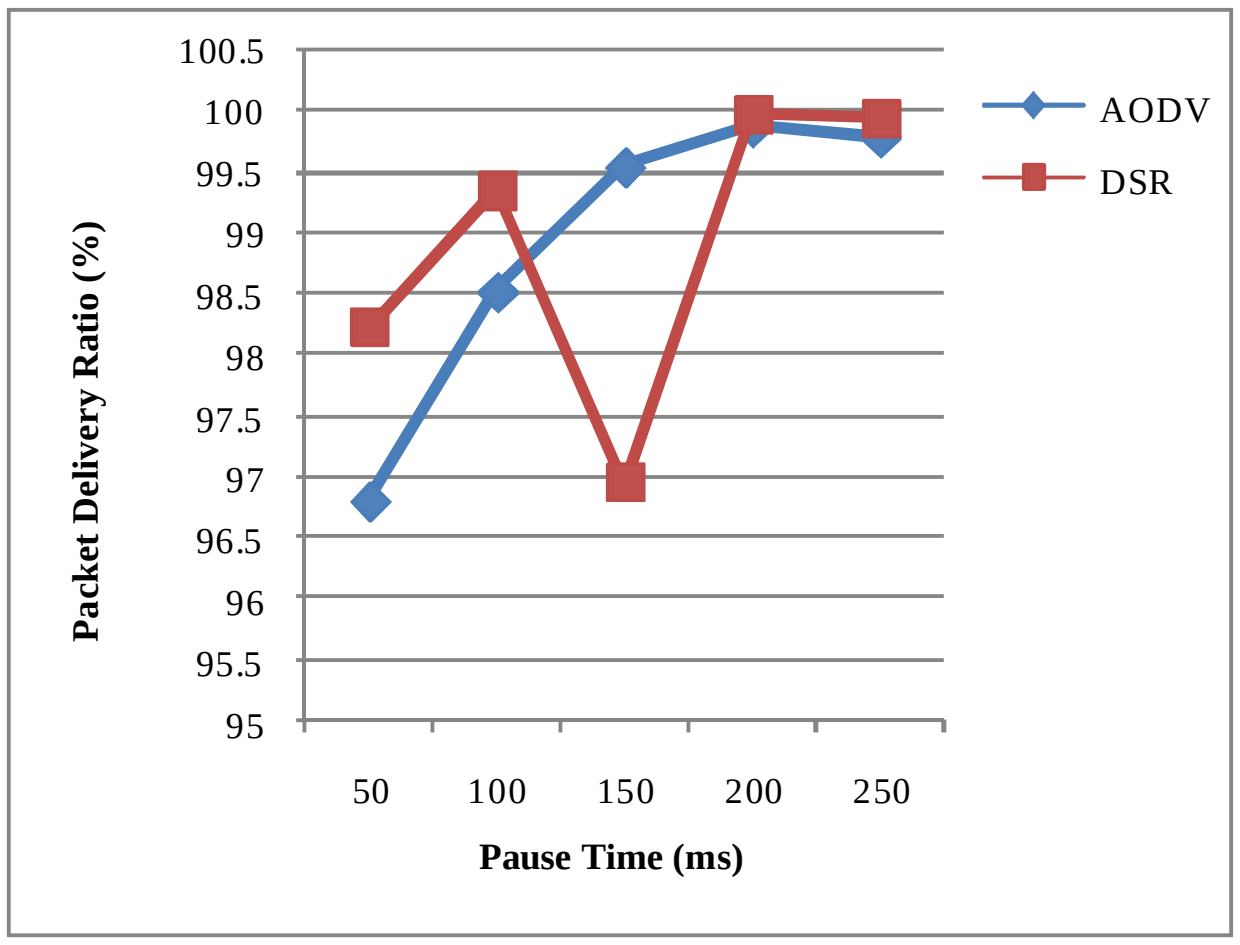

Figure 16. PDR (w.r.t. Pause) of 90 nodes using CBR

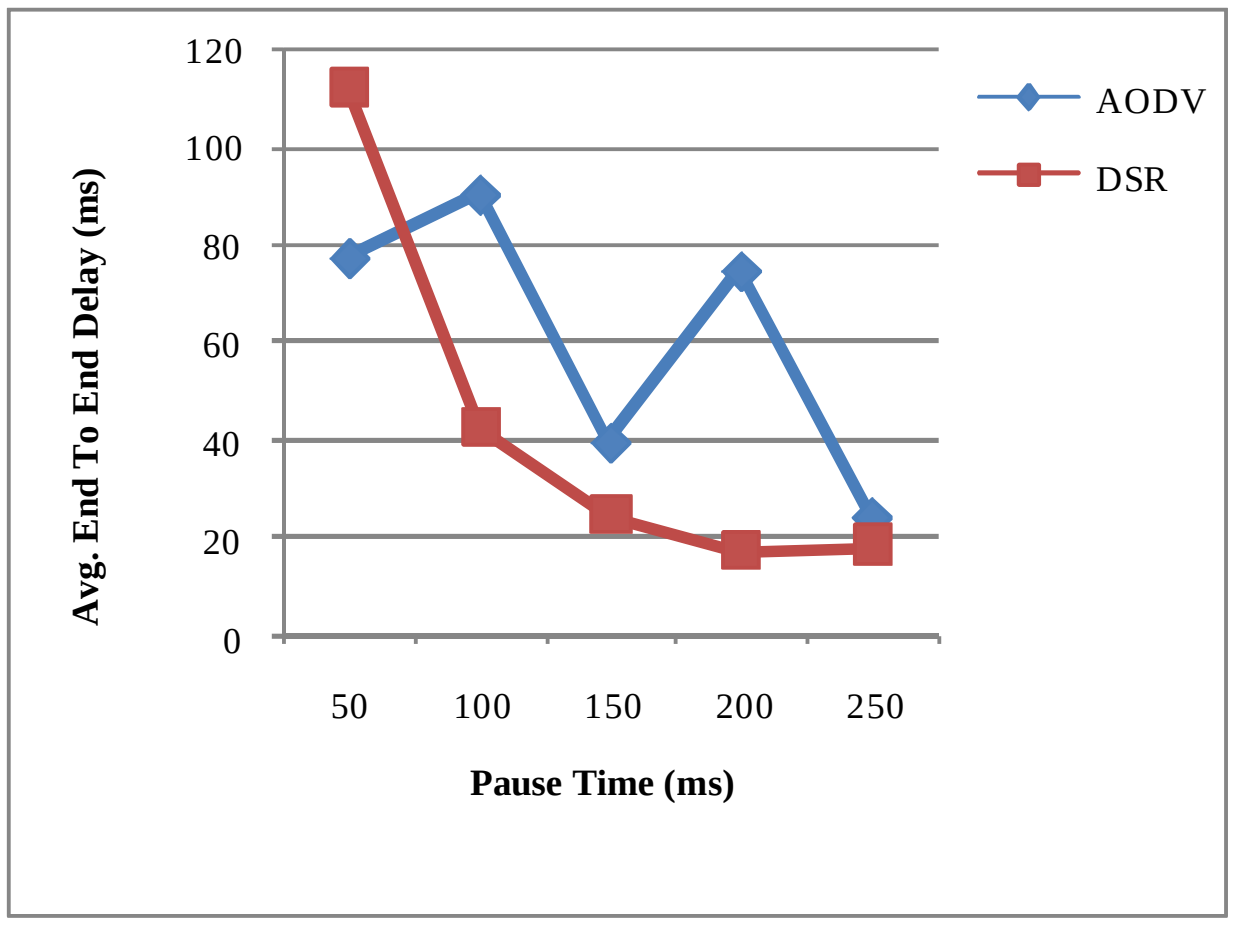

Figure 17. Avg.E-2-E delay (w.r.t. Pause) of 90 nodes using CBR

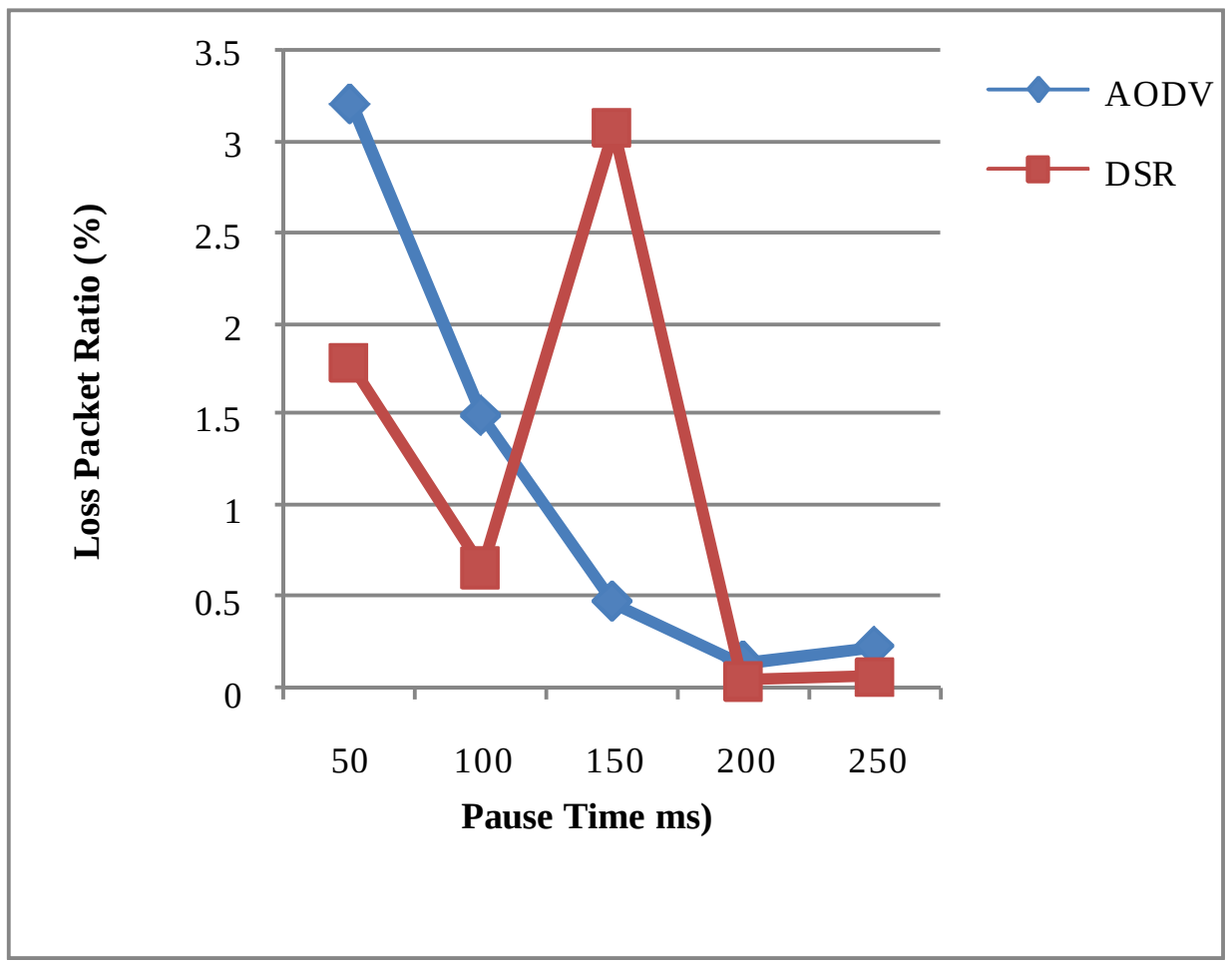

Figure 18. LPR (w.r.t. Pause) of 90 nodes using CBR

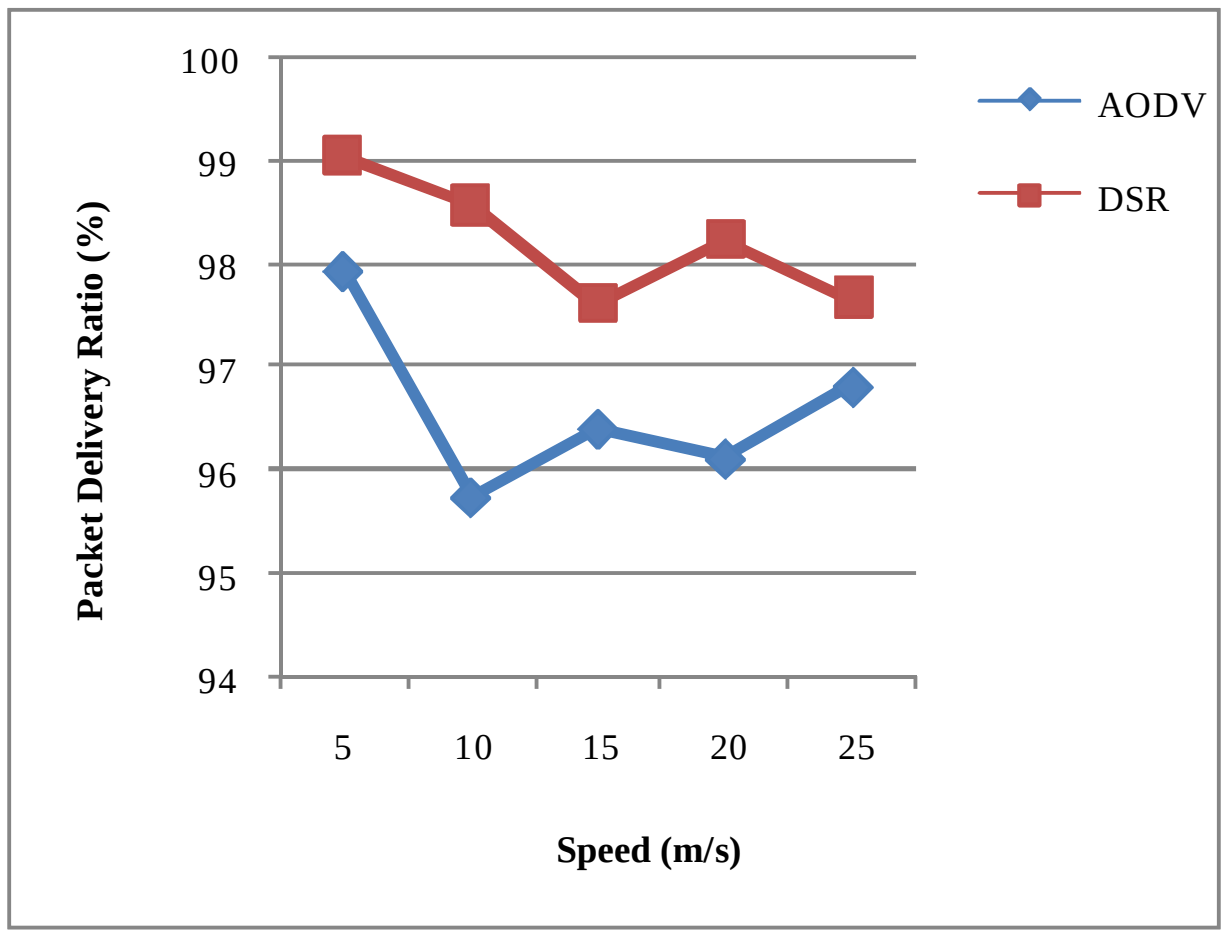

Figure 19. PDR (w.r.t. Speed) of 90 nodes using TCP

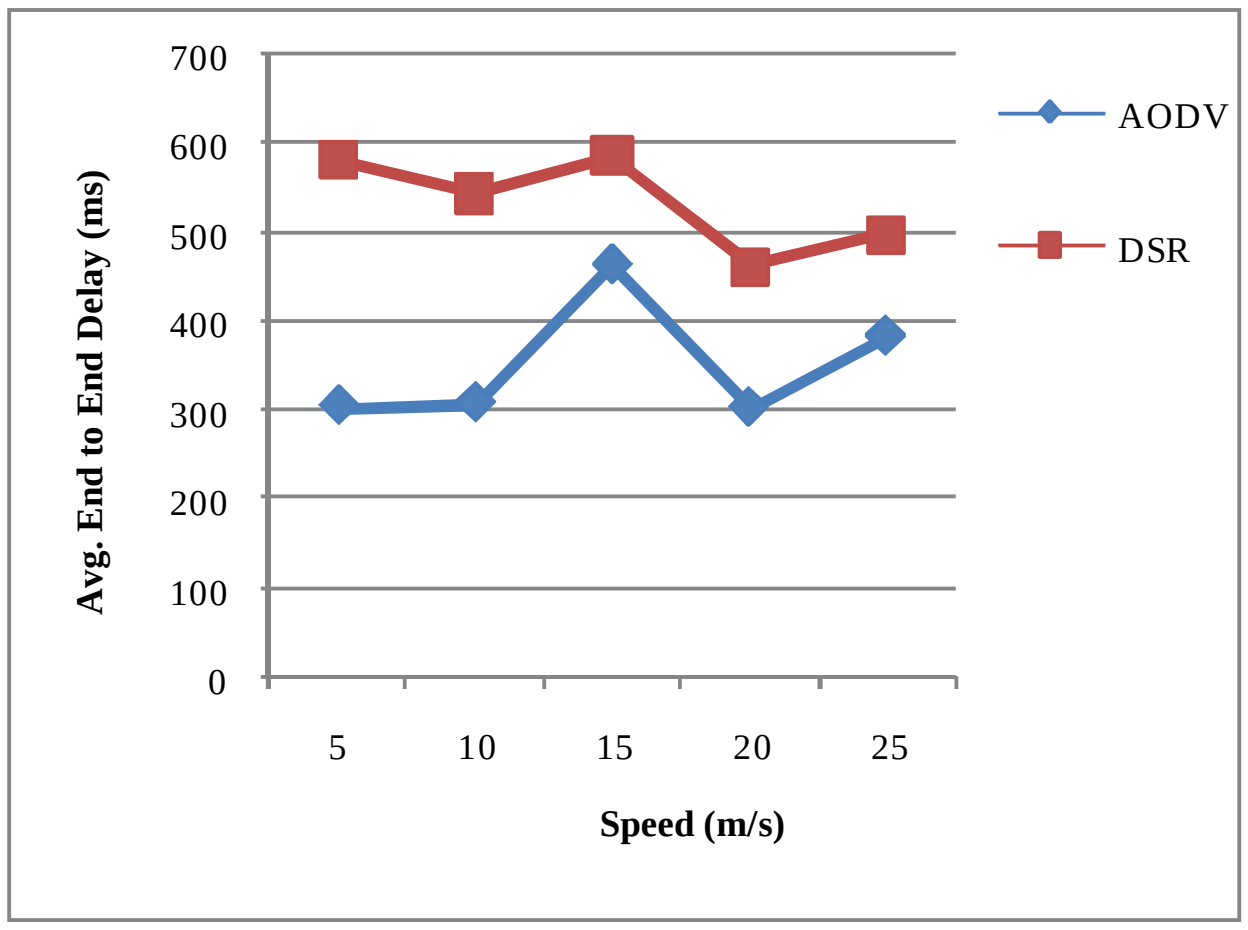

Figure 20. Avg.E2E delay (w.r.t. Speed) of 90 nodes using TCP

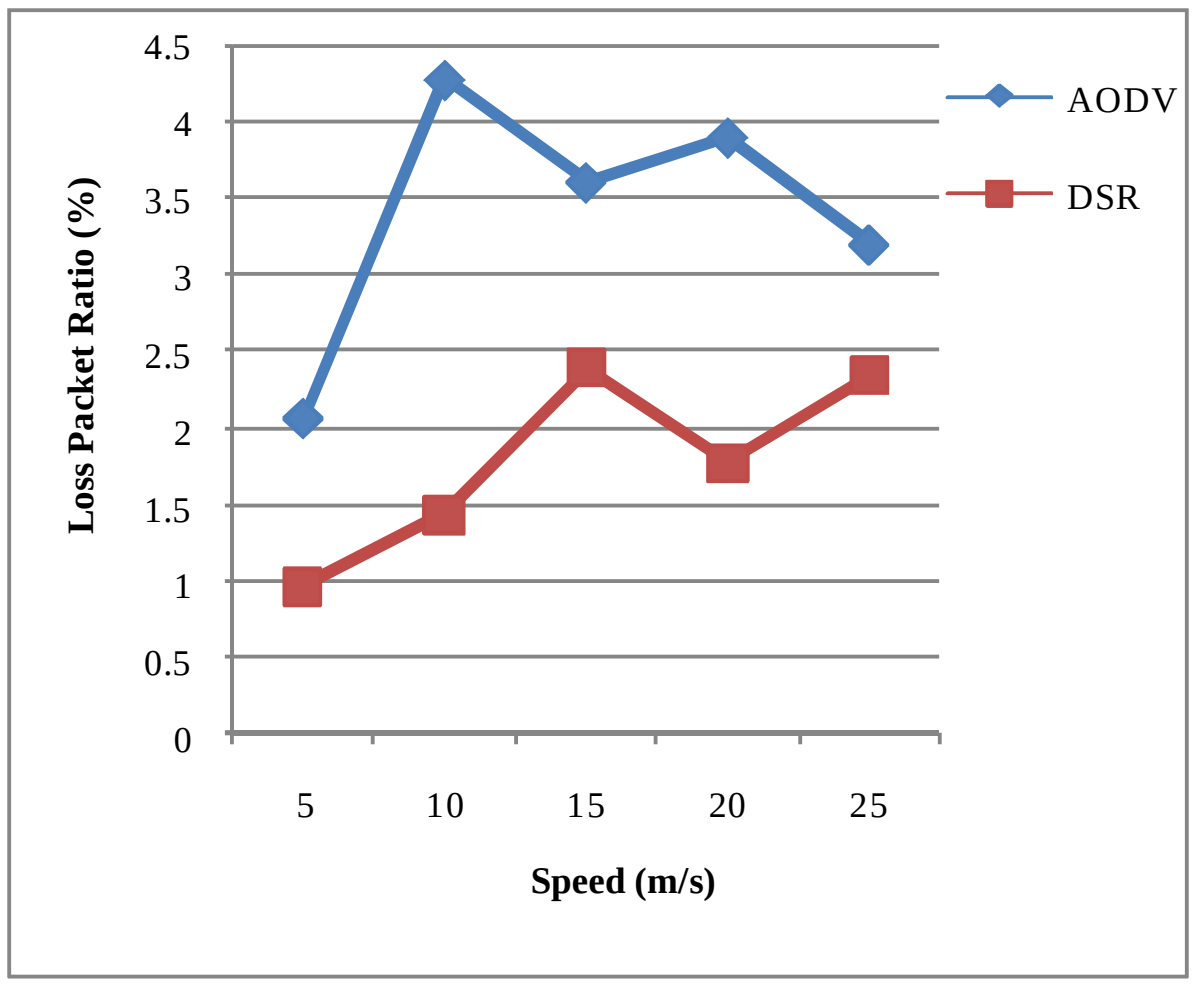

Figure 21. LPR (w.r.t. Speed) of 90 nodes using TCP

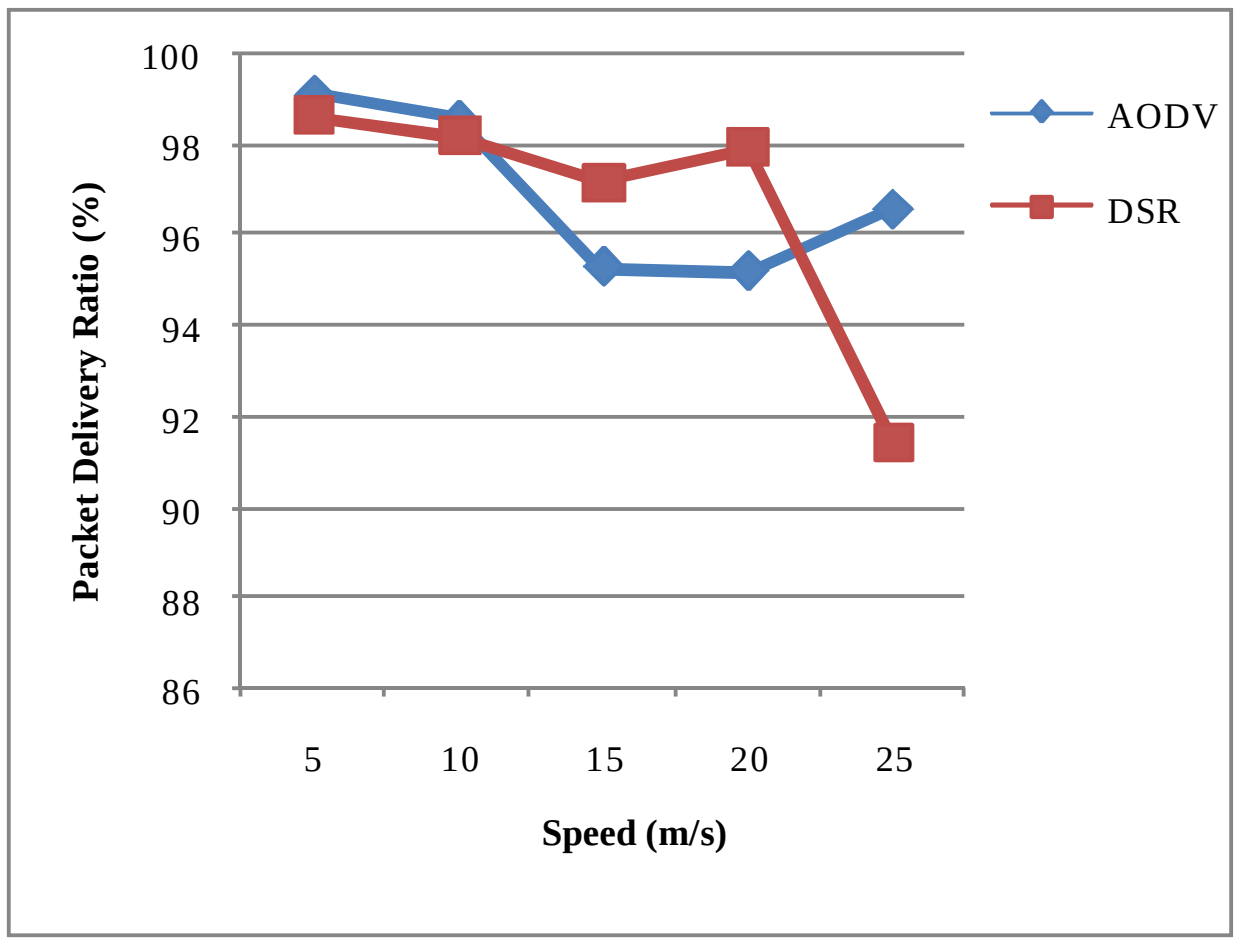

Figure 22. PDR (w.r.t. Speed) of 90 nodes using CBR

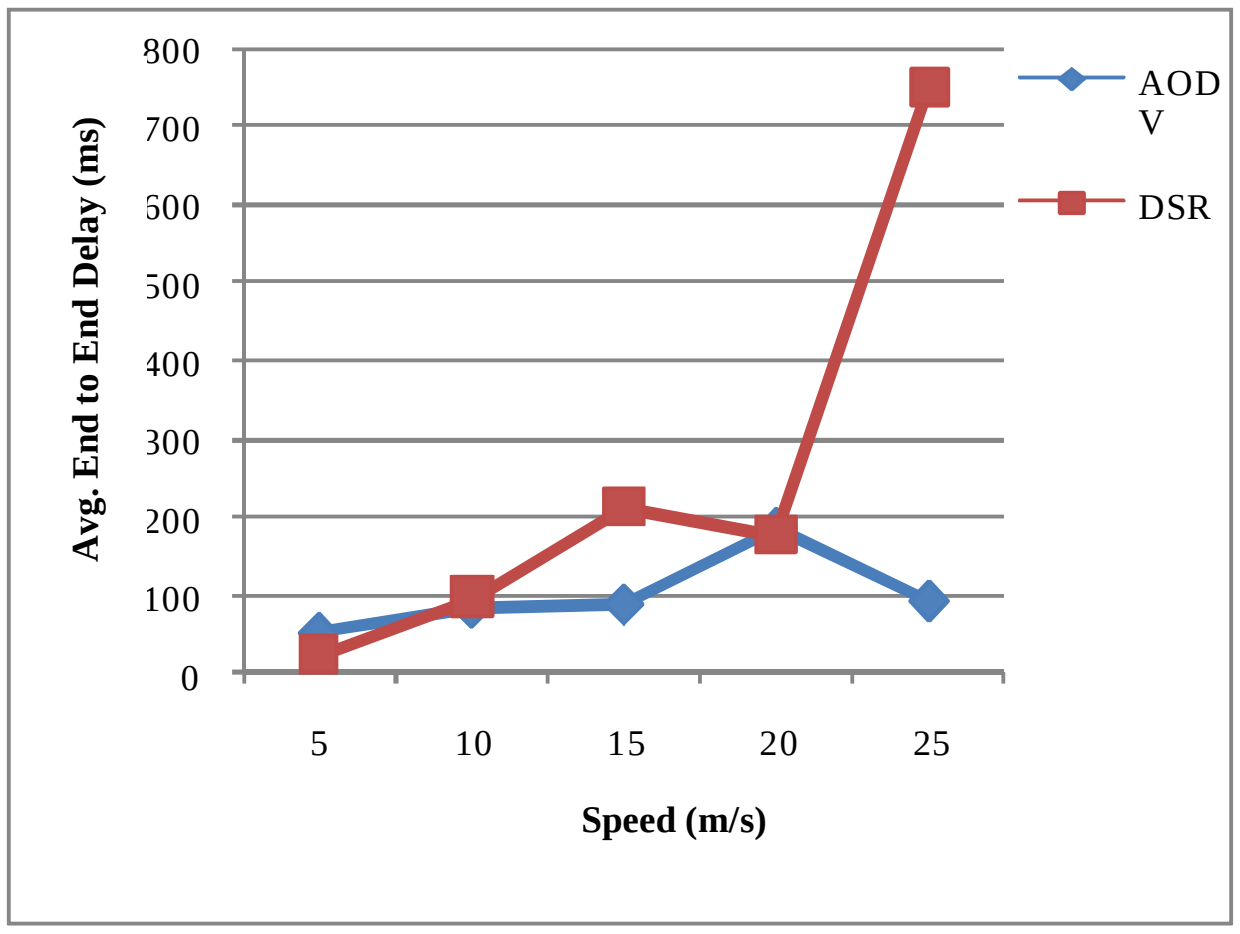

Figure 23. Avg.E2E delay (w.r.t. Speed) of 90 nodes using CBR

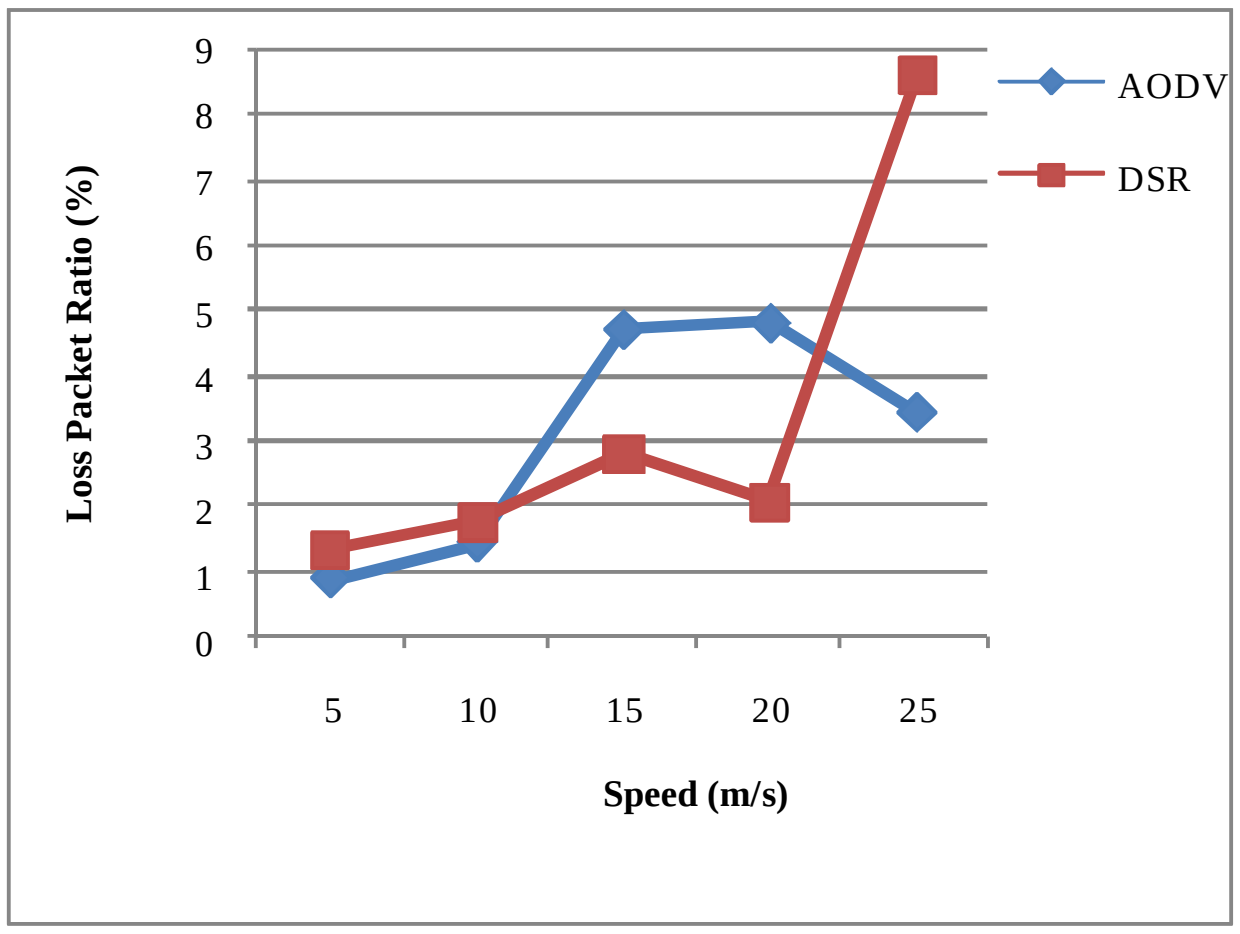

Figure 24. LPR (w.r.t. Speed) of 90 nodes using CBR

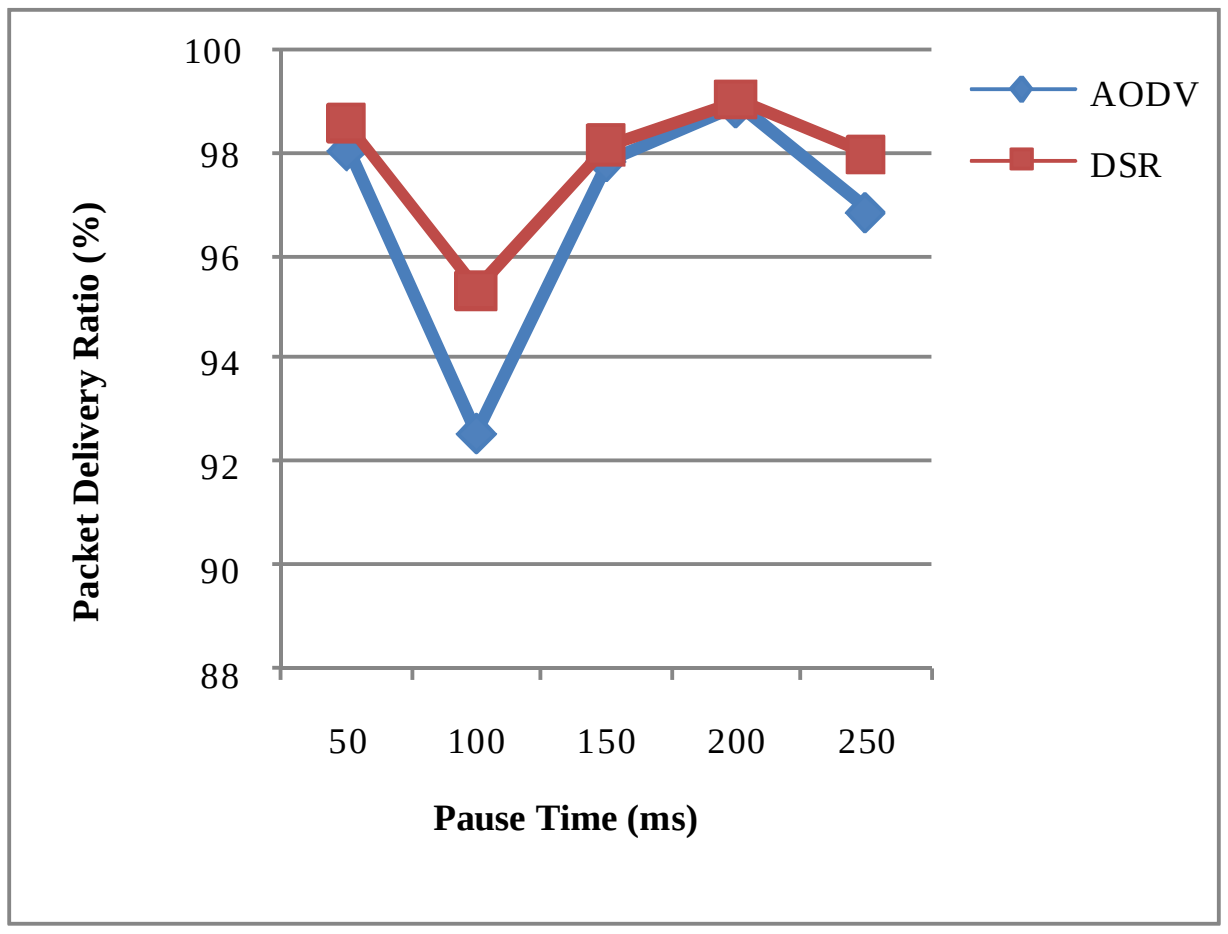

Figure 25. PDR (w.r.t. Pause) of 150 nodes using TCP

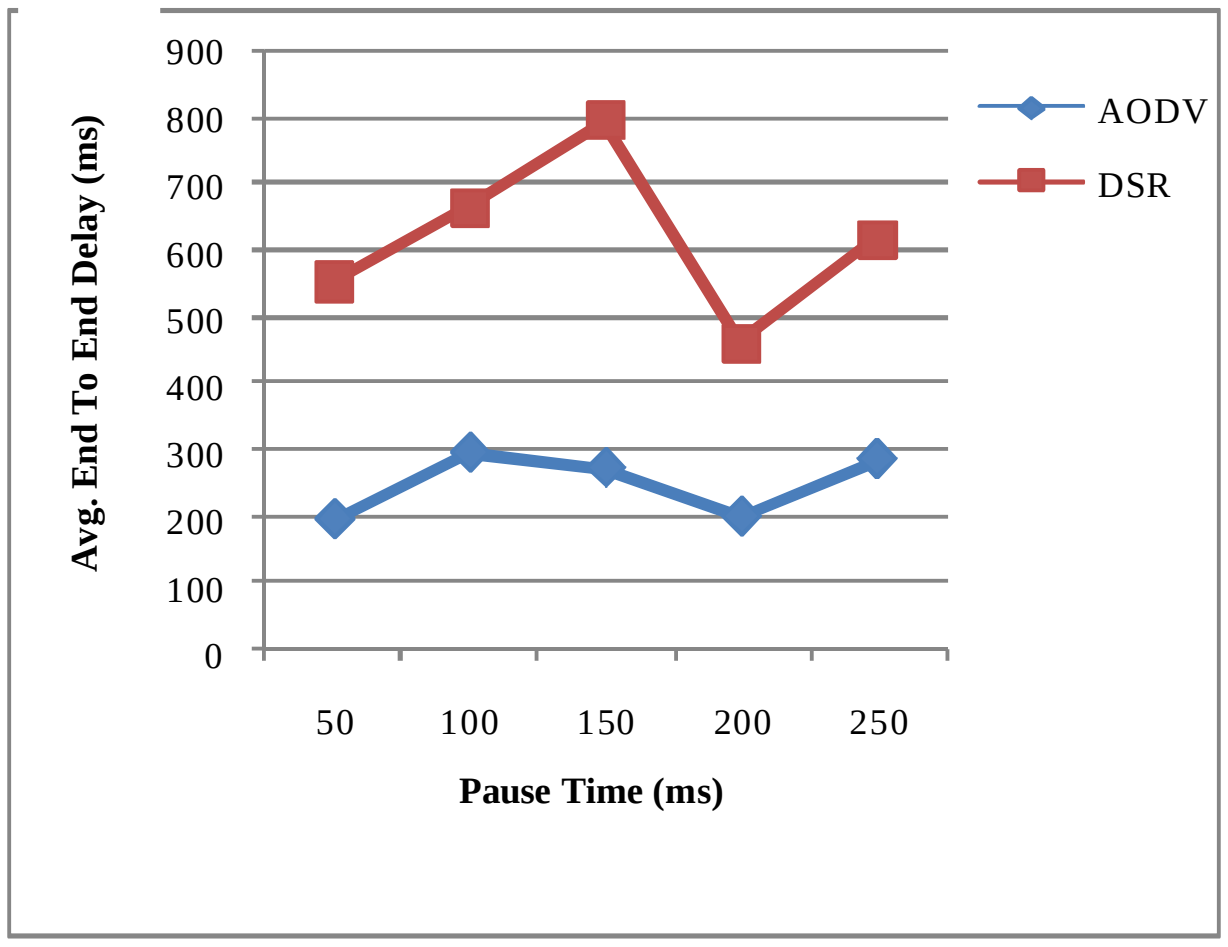

Figure 26. Avg.E2E delay (w.r.t. Pause) of 150 nodes using TCP

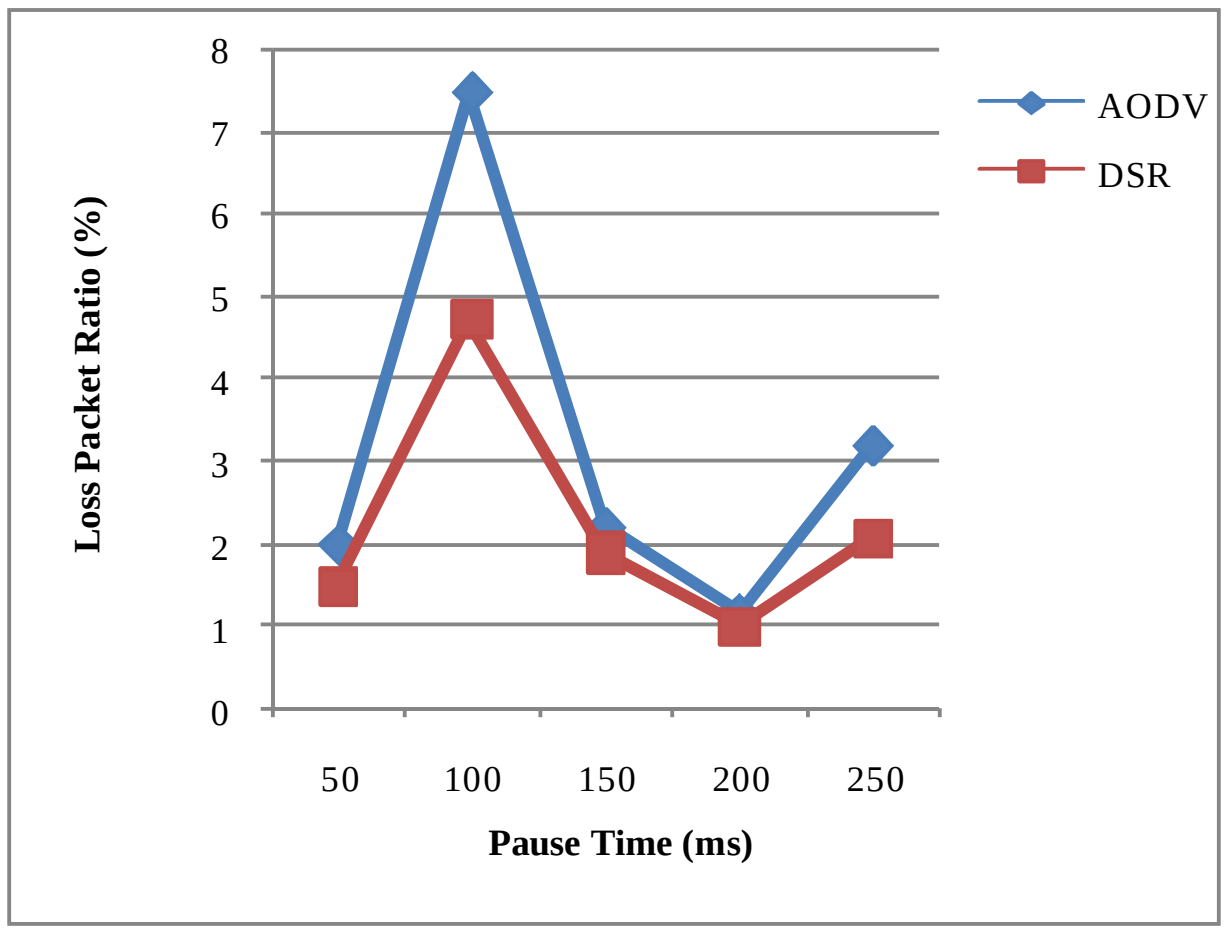

Figure 27. LPR (w.r.t. Pause) of 150 nodes using TCP

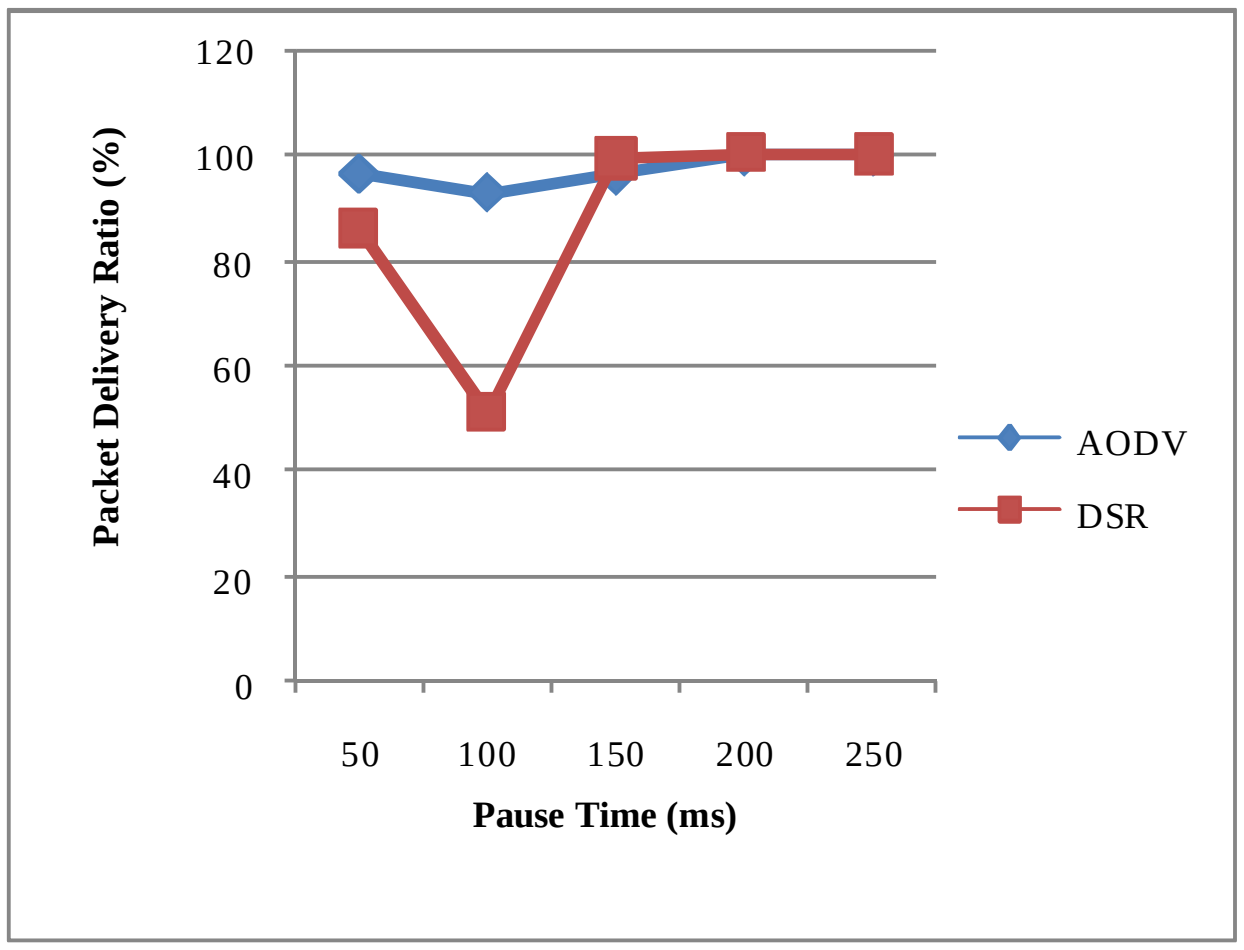

Figure 28. PDR (w.r.t. Pause) of 150 nodes using CBR

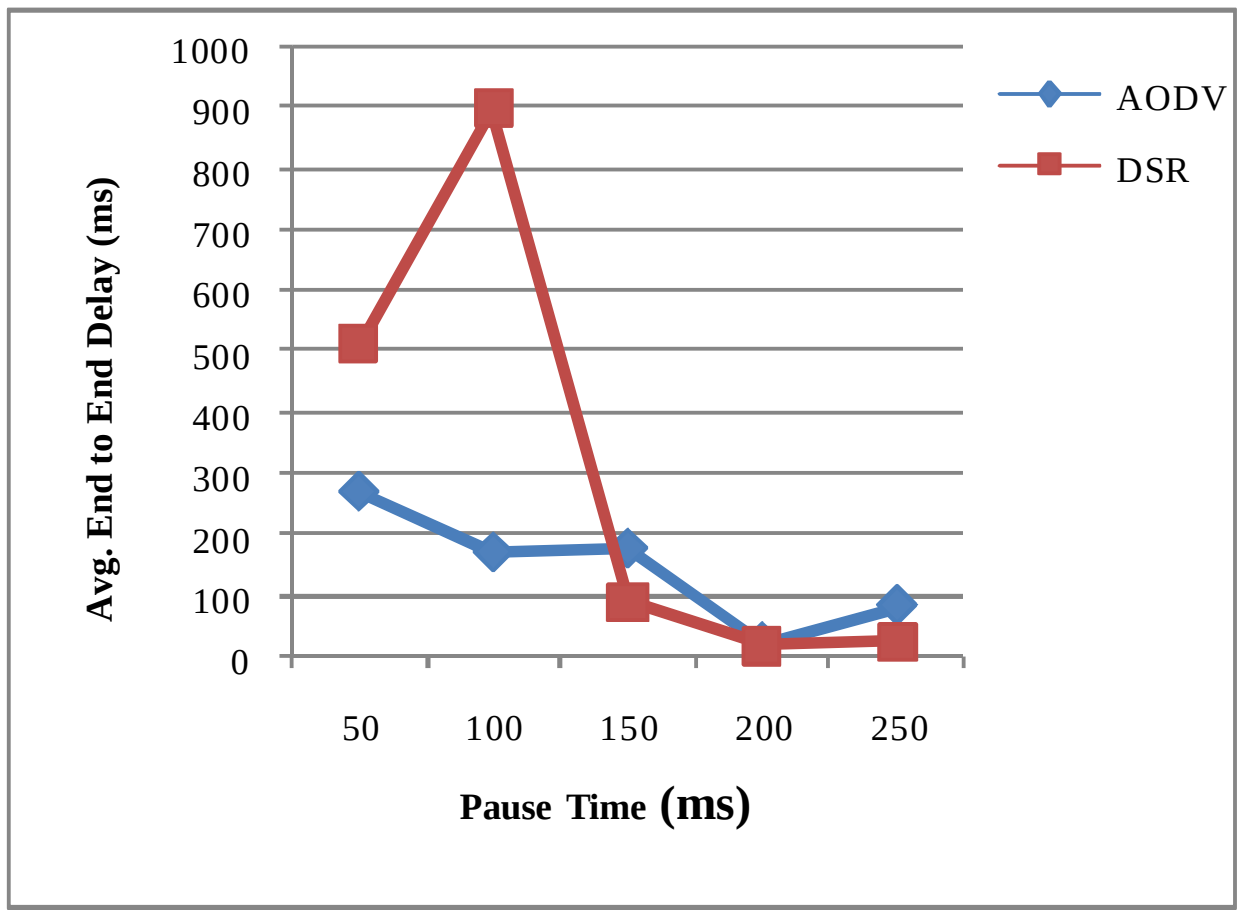

Figure 29. Avg.E2E delay (w.r.t. Pause) of 150 nodes using CBR

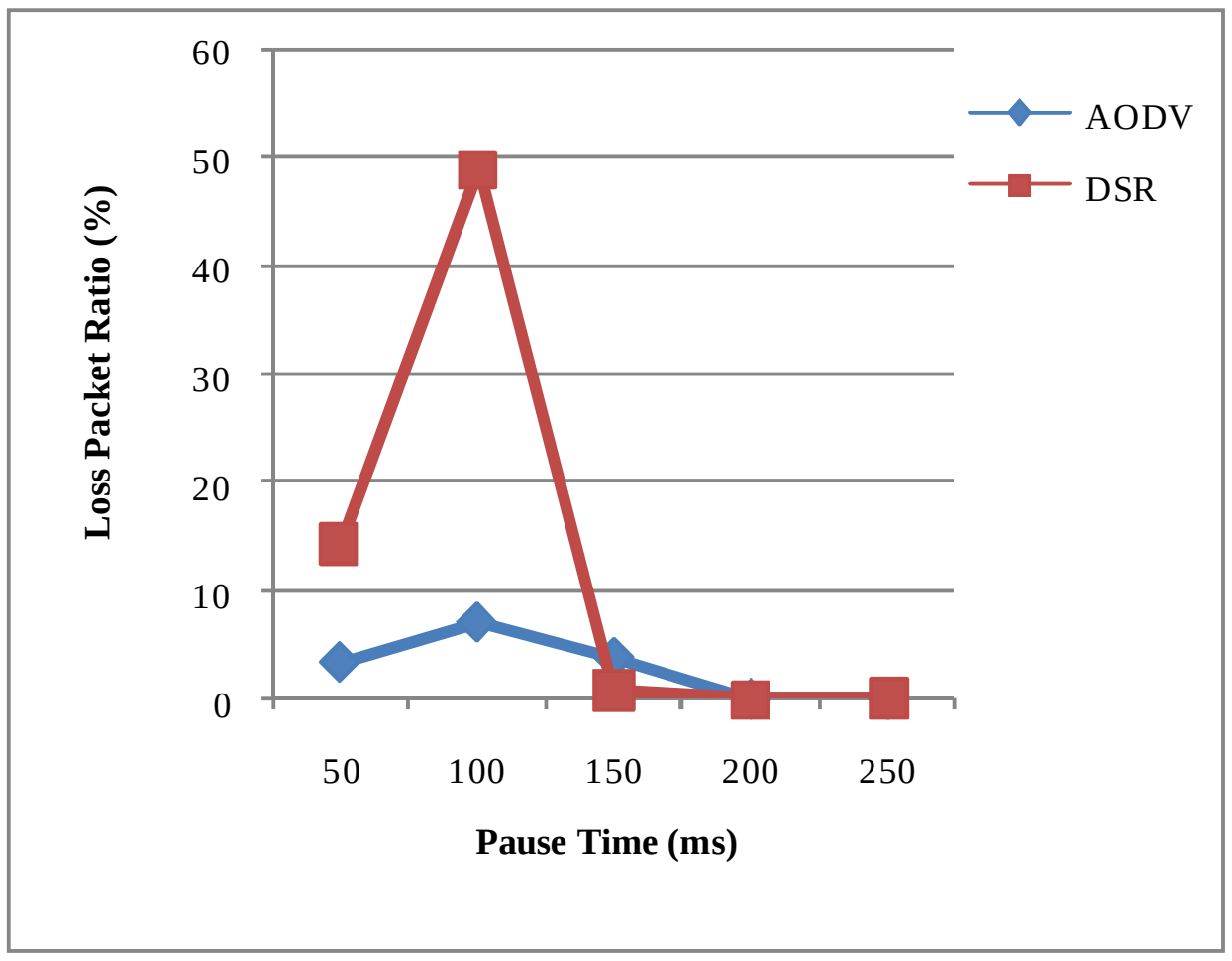

Figure 30. LPR (w.r.t. Pause) of 150 nodes using CBR

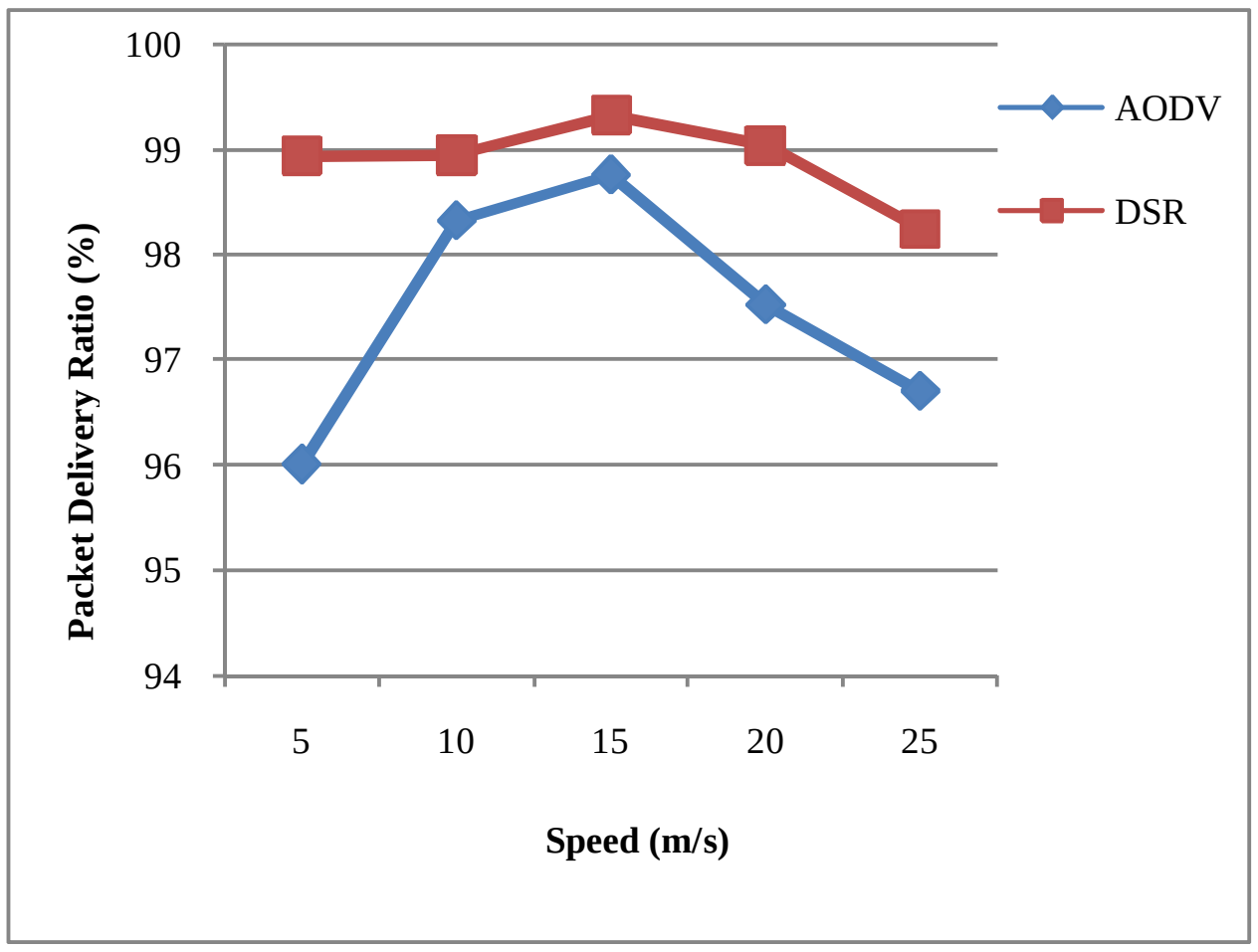

Figure 31: PDR (w.r.t. Speed) of 150 nodes using TCP

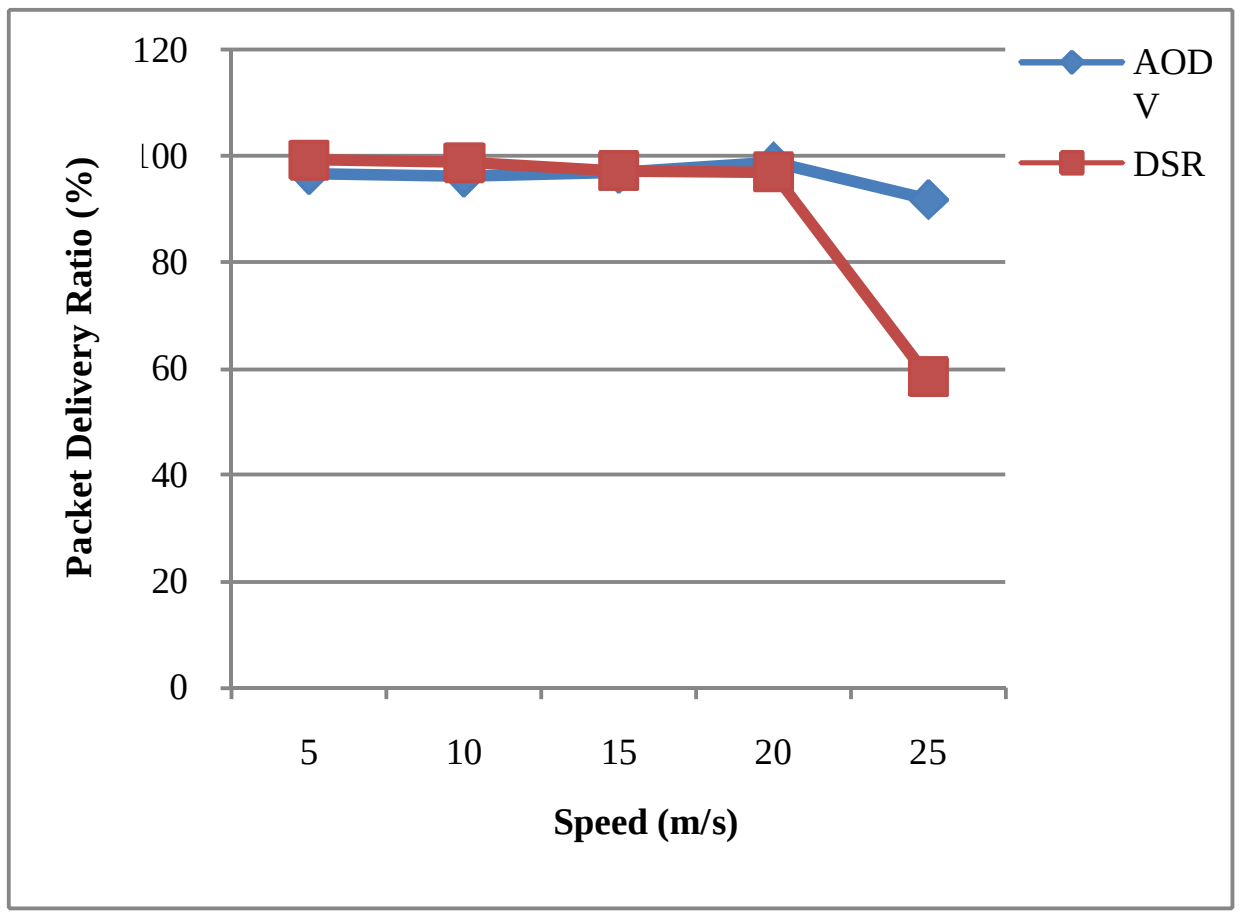

Figure 34: PDR (w.r.t. Speed) of 150 nodes using CBR

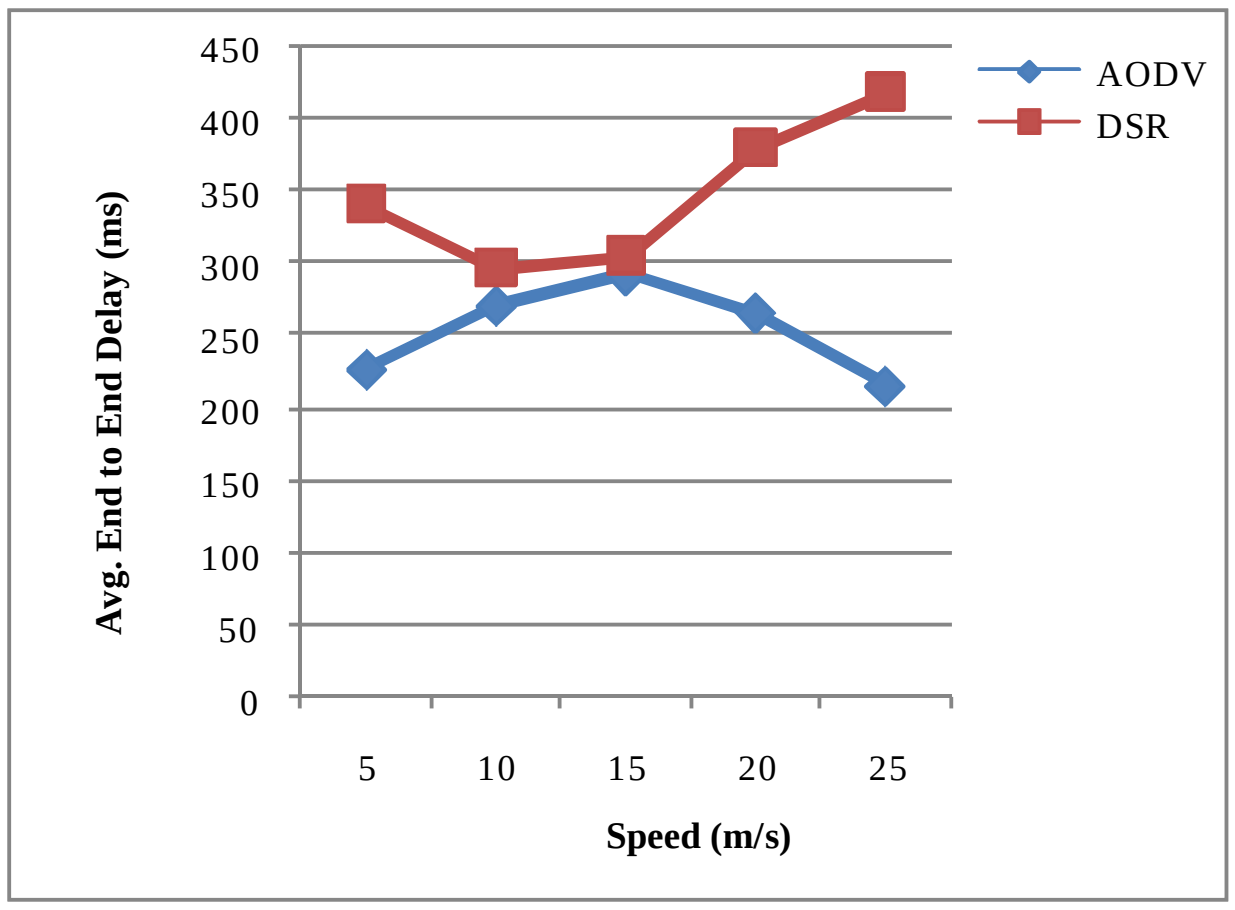

Figure 32: Avg.E2E delay (w.r.t. Speed) of 150 nodes using TCP

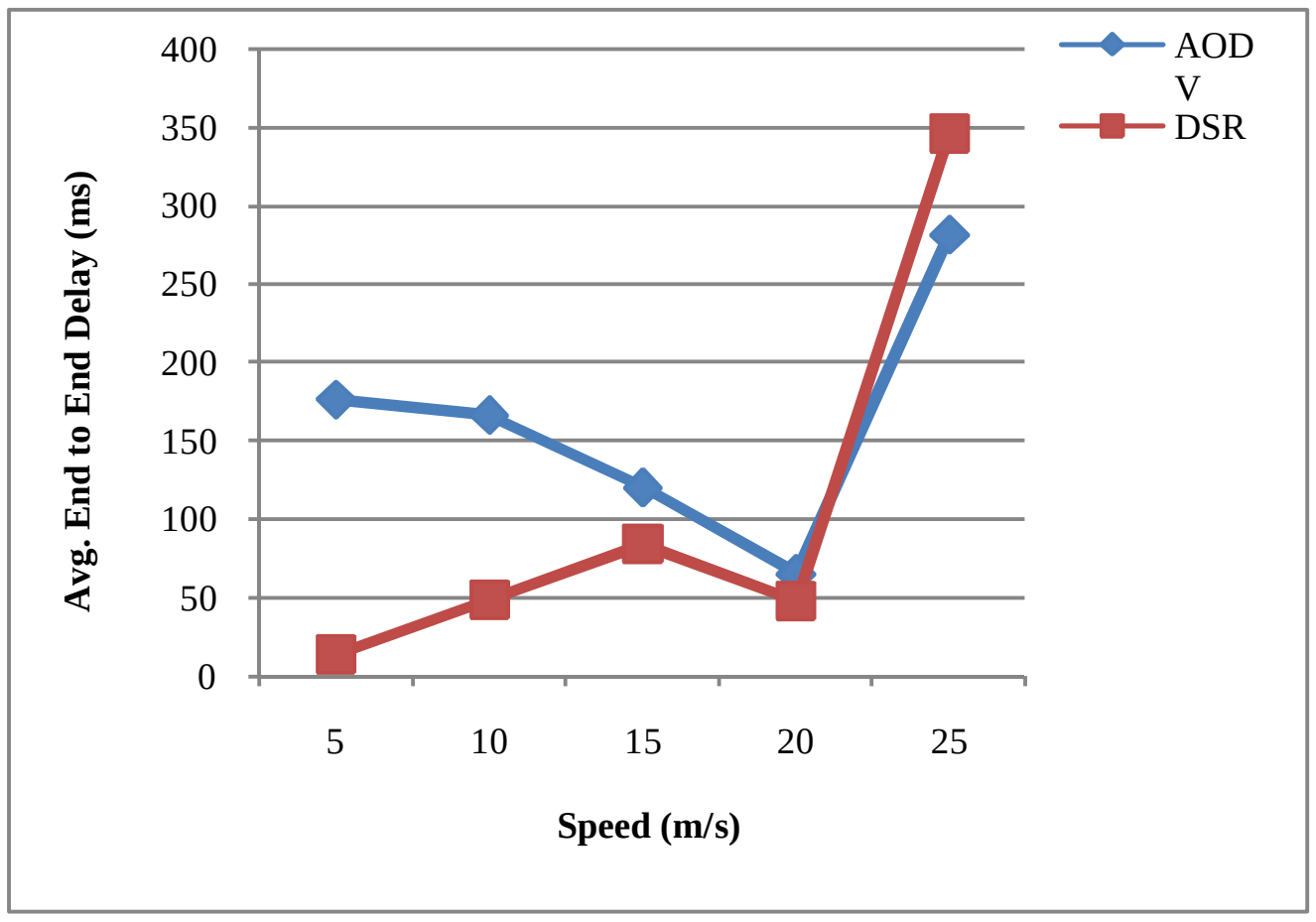

Figure 35: Avg.E2E delay (w.r.t. Speed) of 150 nodes using CBR

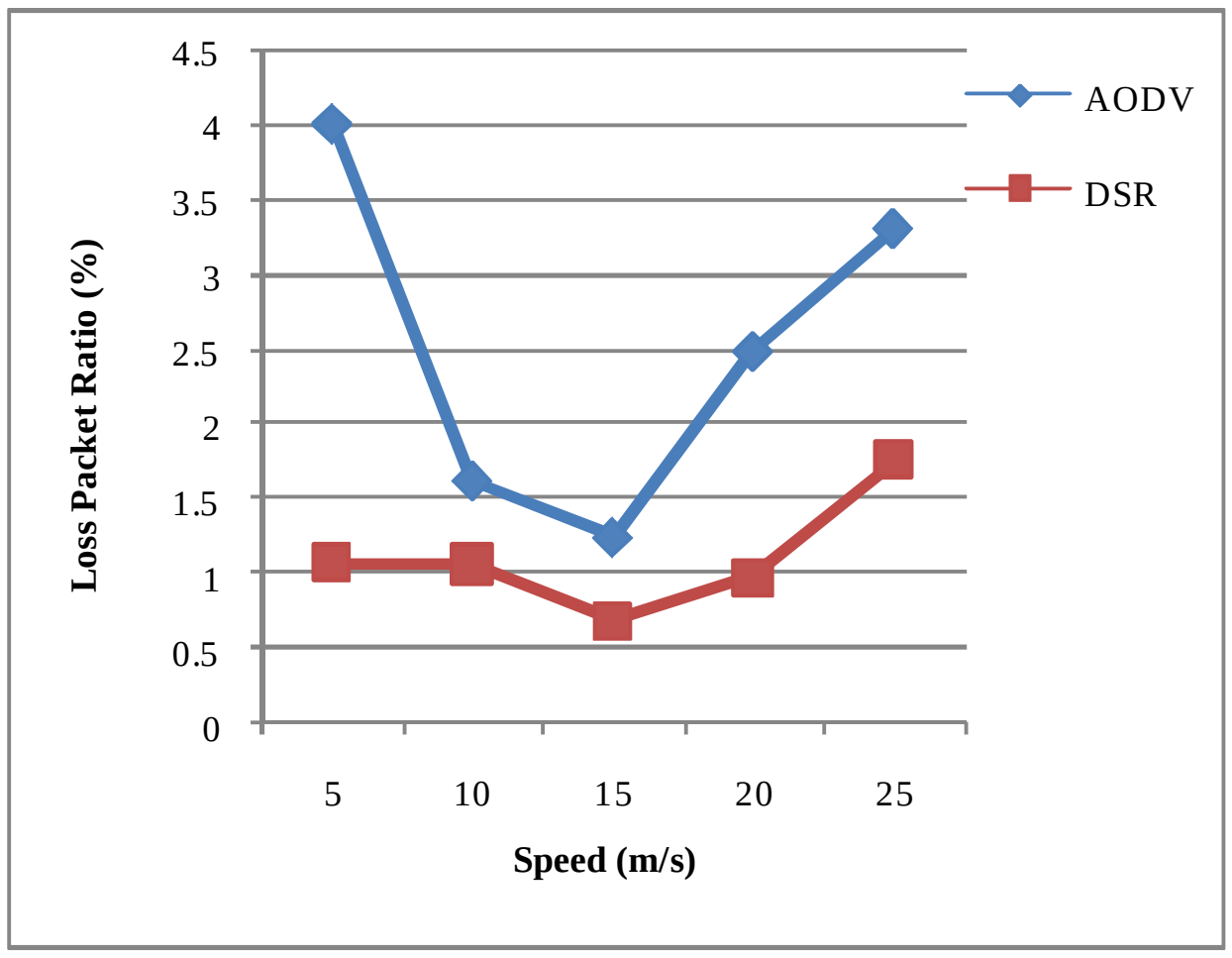

Figure 33: LPR (w.r.t. Speed) of 150 nodes using TCP

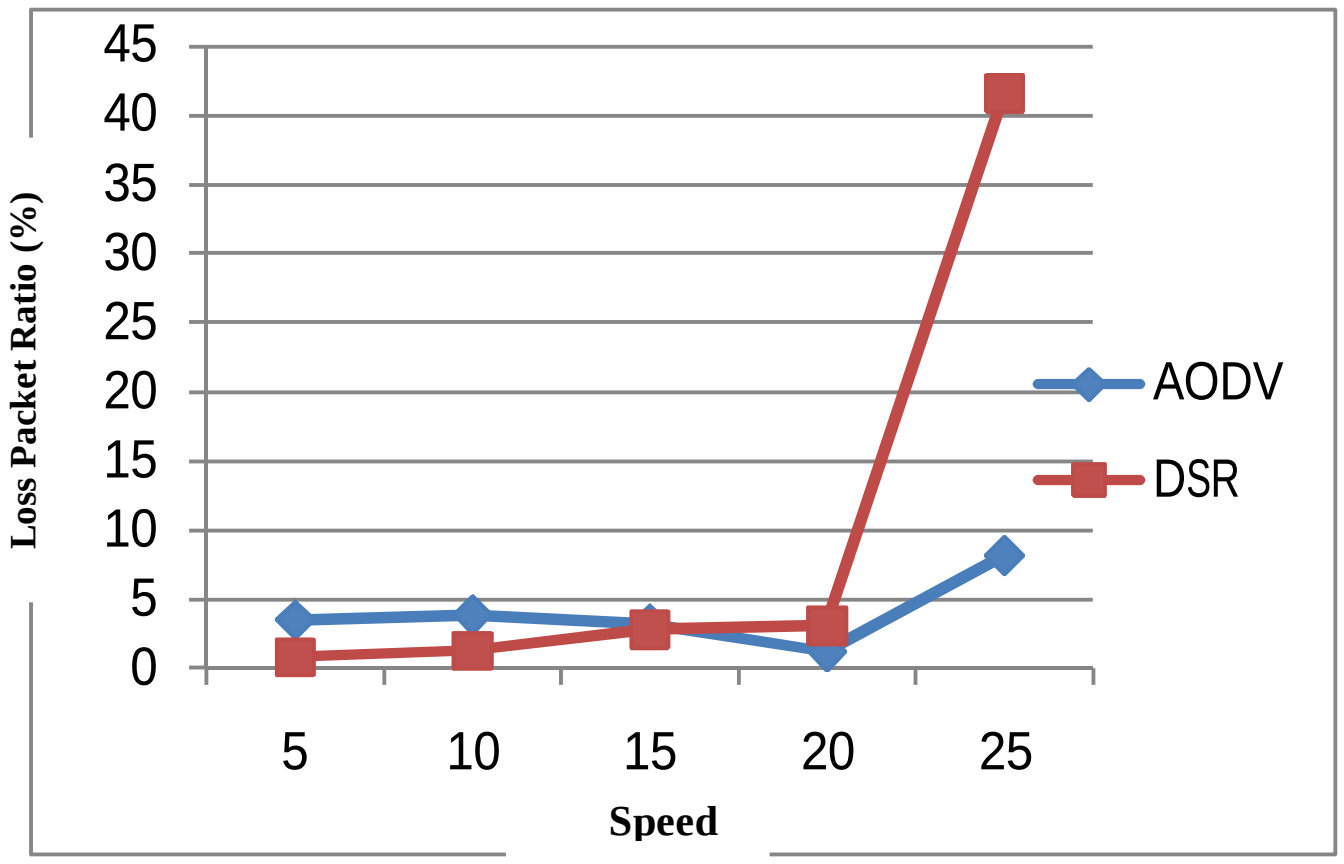

Figure 36: LPR (w.r.t. Speed) of 150 nodes using CBR

6.5 Analysis Table

After analysis of AODV and DSR we define a standard for simulation results. We consider 30 nodes as low density, 90 nodes as average density and 150 nodes as high density. We also consider 5 m/s as low speed, 15 m/s as average speed and 25 m/s as high speed.

The standard for PDR values (approx.) defines below

**For speed & pause time:**

High: >=98%

Average: 96% to 97%

Low: <=95%

The standard for E-to-E values (approx.) defines below

**For pause time:**

High: >=351ms

Average: 151ms to 350ms

Low: <=150ms

**For speed time:**

High: >=150%

Average: 51% to 150%

Low: <=50%

The standard for LPR values (approx.) define below

**For pause time:**

High: > 2%

Average: 1% to 2%

Low: < 1%

**For speed time:**

High: > 3%

Average: 1.5% to 3%

Low: < 1.5%

Our simulation area considered is 840 × 840 and simulation run time is 200 seconds. Speed has been varied from 5m/s to 25 m/s. Pause time has been varied from 50s to 250s. Based on our standard we can summarize the following differences between AODV and DSR based on our estimated parameters.

***Pattern analysis of 30 nodes using TCP connection***

From our experimental analysis we observe that for TCP connection using pause time as a parameter in low mobility low pause time the packet delivery ratio (PDR) is average for AODV and high for DSR. In that scenario average end to end delay (E-To-E) is average for AODV and high for DSR .The loss packet ratio for TCP connection is high for AODV and average for DSR. If the pause time is high the PDR for both routing protocols is high .E-To-E for both protocols is high. LPR of DSR is low but for AODV it is average.

On the other hand, using speed as a parameter in low mobility low speed the packet delivery ratio for both protocols is high. In that scenario average end to end delay (E-To-E) is high, the loss packet ratio is low for both routing protocol. But in low mobility high speed, the PDR for AODV is average but high for DSR. E-To-E for both protocols is high. LPR of AODV is average. But for DSR it is low.

***Pattern analysis of 30 nodes using CBR connection***

We observe that for CBR connection using pause time as a parameter in low mobility low pause time the packet delivery ratio (PDR) of CBR for both routing protocols is high. In that scenario average end to end delay (E-To-E) is low for both protocols .The loss packet ratio is average for AODV and low for DSR. If the pause time is high the PDR for both routing protocols is high .E-To-E is low for both routing protocols. LPR of DSR is low. But for AODV it is low.

On the other hand, using speed as a parameter in low mobility low speed the packet delivery ratio for both protocols is high. In that scenario average end to end delay (E-To-E) and the loss packet ratio is low for both routing protocol. But in low mobility high speed, the PDR for AODV is high but average for DSR. E-To-E for both protocols is low. LPR is average for both routing protocols.

***Pattern analysis of 150 nodes using TCP connection***

Pause time as a parameter in high mobility low pause time PDR for both protocols is high. In that scenario average end to end delay (E-To-E) is average for AODV and high for DSR. The LPR is average for both protocols. If the pause time is high the PDR for both routing protocols is average .E-To-E is average for AODV and high for DSR. LPR is high for AODV and DSR.

On the other hand, using speed as a parameter in high mobility low speed, PDR of AODV is average but high for DSR. Though, E-To-E for AODV & DSR is high. LPR is low for DSR and high for AODV. If the speed is high AODV performs average and DSR performs high .E-To-E is high for both routing protocol. LPR of AODV is high but for DSR it is average.

***Pattern analysis of 150 nodes using CBR connection***

We observe that for CBR connection using pause time as a parameter in high mobility low pause time the packet delivery ratio (PDR) of CBR it is average for AODV and low for DSR. E-To-E for AODV is average but it is high for DSR. The loss packet ratio is high for both protocols. If the pause time is high the PDR for AODV and DSR using CBR is high. .E-To-E and LPR is low for both routing protocols.

On the other hand, using speed as a parameter in high mobility low speed the packet delivery ratio for AODV is average but high for DSR, Though E-To-E and LPR for AODV is high but low for DSR. If the speed is high the PDR for AODV and DSR is low .E-To-E is high for both routing protocol. LPR of AODV and DSR is high for CBR connection.

## 7. OUR DECISIONS

After performance analysis of AODV & DSR by using decision table we declare our decision.

TABLE-2: PDR, E-2-E AND LPR WITH RESPECT TO LOW MOBILITY & LOW PAUSE TIME FOR TCP & CBR CONNECTIONS

| Protocols | Packet Delivery Ratio | | Avg. End to End Delay | | Loss Packet Ratio | |
|---|---|---|---|---|---|---|
| | ***TCP*** | ***CBR*** | ***TCP*** | ***CBR*** | ***TCP*** | ***CBR*** |
| AODV | Avg | High | Avg | Low | High | Avg |
| DSR | High | High | High | Low | Avg | Low |

TABLE-3: PDR, E-2-E AND LPR WITH RESPECT TO LOW MOBILITY & HIGH PAUSE TIME FOR TCP & CBR CONNECTIONS

| Protocols | Packet Delivery Ratio | | Avg. End to End Delay | | Loss Packet Ratio | |
|---|---|---|---|---|---|---|
| | ***TCP*** | ***CBR*** | ***TCP*** | ***CBR*** | ***TCP*** | ***CBR*** |
| AODV | High | High | High | Low | Avg | Low |
| DSR | High | High | High | Low | Low | Low |

TABLE-4: PDR, E-2-E AND LPR WITH RESPECT TO LOW MOBILITY & LOW SPEED TIME FOR TCP & CBR CONNECTIONS

| Protocols | Packet Delivery Ratio | | Avg. End to End Delay | | Loss Packet Ratio | |
|---|---|---|---|---|---|---|
| | ***TCP*** | ***CBR*** | ***TCP*** | ***CBR*** | ***TCP*** | ***CBR*** |
| AODV | High | High | High | Low | Low | Low |
| DSR | High | High | High | Low | Low | Low |

TABLE-5:PDR, E-2-E AND LPR WITH RESPECT TO LOW MOBILITY & HIGH SPEED TIME FOR TCP & CBR CONNECTIONS

| Protocols | Packet Delivery Ratio | | Avg. End to End Delay | | Loss Packet Ratio | |
|---|---|---|---|---|---|---|
| | ***TCP*** | ***CBR*** | ***TCP*** | ***CBR*** | ***TCP*** | ***CBR*** |
| AODV | Avg | High | High | Avg | Avg | Avg |
| DSR | High | Avg | High | Avg | Low | Avg |

TABLE-6: PDR, E-2-E AND LPR WITH RESPECT TO HIGH MOBILITY & LOW PAUSE TIME FOR TCP & CBR CONNECTIONS

| Protocols | Packet Delivery Ratio | | Avg. End to End Delay | | Loss Packet Ratio | |
|---|---|---|---|---|---|---|
| | ***TCP*** | ***CBR*** | ***TCP*** | ***CBR*** | ***TCP*** | ***CBR*** |
| AODV | High | Avg | Avg | Avg | Avg | High |
| DSR | High | Low | High | High | Avg | High |

TABLE-7:PDR E-2-E AND LPR WITH RESPECT TO HIGH MOBILITY & HIGH PAUSE TIME FOR TCP & CBR CONNECTIONS

| Protocols | Packet Delivery Ratio | | Avg. End to End Delay | | Loss Packet Ratio | |
|---|---|---|---|---|---|---|
| | ***TCP*** | ***CBR*** | ***TCP*** | ***CBR*** | ***TCP*** | ***CBR*** |
| AODV | Avg | High | Avg | Low | High | Low |
| DSR | Avg | High | High | Low | High | Low |

TABLE-8: PDR, E-2-E AND LPR WITH RESPECT TO HIGH MOBILITY & LOW SPEED TIME FOR TCP & CBR CONNECTIONS

| Protocols | Packet Delivery Ratio | | Avg. End to End Delay | | Loss Packet Ratio | |
|---|---|---|---|---|---|---|
| | ***TCP*** | ***CBR*** | ***TCP*** | ***CBR*** | ***TCP*** | ***CBR*** |
| AODV | Avg | Avg | High | High | High | High |
| DSR | High | High | High | Low | Low | Low |

TABLE-9:PDR, E-2-E AND LPR WITH RESPECT TO HIGH MOBILITY & HIGH SPEED TIME FOR TCP & CBR CONNECTIONS

| Protocols | Packet Delivery Ratio | | Avg. End to End Delay | | Loss Packet Ratio | |
|---|---|---|---|---|---|---|
| | ***TCP*** | ***CBR*** | ***TCP*** | ***CBR*** | ***TCP*** | ***CBR*** |
| AODV | Avg | Low | High | High | High | High |
| DSR | High | Low | High | High | Avg | High |

## 8.CONCLUSION

In the research paper we mainly analysis the performance of two on demand routing protocols AODV and DSR on the basis of packet delivery ratio, average End-to-End delay and Loss packet ratio. We observe that the performance of AODV and DSR depends on scenario. The performance measurement of AODV and DSR will help for further development of these protocols in future.

**Bijan Paul** is a B.Sc. student in the Dept. of Computer Science & Engineering, Shahjalal University of Science & Technology, Bangladesh. His research interests include VANET, Routing protocols, Wireless computing.

**Md. Ibrahim** is a B. Sc. student in the Dept. of Computer Science & Engineering, Shahjalal University of Science & Technology, Bangladesh. His research interest includes VANET, Routing protocols, Wireless Computing.

**Md. Abu Naser Bikas** obtained his B. Sc. Degree in Computer Science & Engineering from Shahjalal University of Science & Technology, Bangladesh. Currently, he is a Lecturer in Computer Science & Engineering at the same University. His research interests include VANET, Network Security, Intrusion Detection and Intrusion Prevention, Bangla OCR, Wireless Ad-Hoc Networks, and Grid Computing.